\documentclass[a4paper]{aa}

\usepackage{graphicx}
\usepackage{txfonts}
\usepackage{lscape}
\usepackage{tabularx}
\usepackage{adjustbox}
\usepackage{hyperref}
\usepackage{comment}
\usepackage{stfloats}

\begin{document} 
   \title{Chemical complexity in feedback from supernova remnants:\\ first detection of PO$^+$ in IC443 and W44.}

\author{G. Cosentino$^{1}$\thanks{E-mail:cosentino@iram.fr},
F. D. Priestley$^2$,
S. Scibelli$^3$\thanks{Jansky Fellow of the National Radio Astronomy Observatory.},
F. Fontani$^{4,5,6}$
}
\authorrunning{Cosentino et al.}
    \titlerunning{Chemical complexity in SNRs}
    \institute{Institut de Radioastronomie Millimétrique, 300 Rue de la Piscine, 38400 Saint-Martin-d’Hères, France
    \and School of Physics and Astronomy, Cardiff University, Queen’s Buildings, The Parade, Cardiff CF24 3AA, UK
    \and National Radio Astronomy Observatory, 520 Edgemont Road, Charlottesville, VA 22903, USA
    \and INAF  Osservatorio Astronomico di Arcetri, Largo E. Fermi 5, 50125 Florence, Italy
    \and Max Planck Institute for Extraterrestrial Physics, Giessenbachstrasse 1, 85748 Garching bei M\"{u}nchen, Germany
    \and Laboratory for the study of the Universe and eXtreme phenomena (LUX), Observatoire de Paris, 5, place Jules Janssen, 92195 Meudon, France}

   \date{Received ---; accepted ---}

   \date{Received XX, XXXX; accepted March XX, XXXX}

% \abstract{}{}{}{}{} 
% 5 {} token are mandatory
 
  \abstract
   {Supernova remnant (SNR) shocks may shape the initial physical and chemical properties of star formation in galaxies. By releasing refractory elements into the gas phase, they may inject material of pre-biotic relevance into molecular clouds. Such elements may later on be inherited by planetary systems and therefore be relevant for the emergence of life as we know it.}
   {Here, we investigate the presence of Phosphorus-bearing (P-bearing) species in molecular clouds impacted by SNR-shocks. Phosphorus is a crucial biological element, but its presence and main reservoir in molecular clouds remain highly elusive.}
   {We present high-sensitivity 3 mm single-pointing observations obtained with the 30 m antenna at the Instituto de Radioastronomia Mill\'imetrica, toward the interaction sites between the SNRs W44 and IC443 and their associated molecular clouds. Focusing on the detection of PO$^+$(2-1) and PN(2-1), we derive molecular column densities and abundances in local thermodynamic equilibrium, discussing the implication of this assumption. We also investigate the correlation between P-bearing species and the shock tracer Silicon Monoxide (SiO).}
   {We detect, for the first time, PO$^+$ emission toward SNRs, with abundances $\chi$(PO$^+$)$\sim$ 8$\times$10$^{-12}$-5$\times$10$^{-11}$. On the other hand, no significant PN emission is detected.  PO$^+$ toward both SNRs is associated with strong SiO emission, showing similar linewidths but systematic velocity offsets, suggesting that PO$^+$ preferentially traces the most compressed and possibly more strongly ionised layers of the shock. Moreover, by comparing the derived abundance with those reported in literature toward other star-forming regions, we suggest that the PO$^+$ abundance mainly depends on a combination of shock strength and shocked medium density.}
   {Our results indicate that phosphorus chemistry in SNR-driven shocks is shifted toward an ion-dominated regime, likely driven by enhanced cosmic-ray ionisation rates. In this scenario, neutral species such as PN are efficiently destroyed, while ion-molecule reactions favour the production of PO$^+$. These findings suggest that SNR-driven shocks may play a role in processing phosphorus-bearing material and shaping gas-phase phosphorus chemistry in molecular clouds, with potential implications for the chemical inventory of star-forming environments.}


   \maketitle
   \nolinenumbers
%
%-------------------------------------------------------------------

\section{Introduction}
In recent years, both theoretical \citep[e.g. ][]{bigiel2008,bigiel2010,leroy2013,hopkins2018} and observational studies \citep[e.g. ][]{cosentino2019,RicoVillas2020,sano2020,sano2023,Reach2024,cosentino2025a,cosentino2026} have indicated feedback from Supernova Remnants (SNRs) as powerful events able to affect the physics and chemistry of star formation in galaxies \citep{chevalier1974,slane2016,Trapp2024}. Through both compressive \citep{Federrath2012,inutsuka2015,Appel2023} and dispersive \citep{kortgen2016,kruijssen2019} mechanisms, SNR-driven shocks have the potential to both trigger and hinder star formation in molecular clouds, i.e. the coldest \citep[T$<$25 K;][]{pillai2006} and densest \citep[n(H$_2$)$>$10$^3$ cm$^{-3}$;][]{butlerTan2012} phase of the interstellar medium (ISM). Direct observational evidence of SNR-triggered star formation has been recently reported by \cite{cosentino2025a}.  

Our current understanding suggests that even our own Sun might have formed throughout this route \citep{Boss2013,Young2014,Parker2023}. In this scenario, a SNR-driven shock impacted the protosolar nebula, triggering its collapse and simultaneously enriching its material with chemically processed elements from the ejecta \citep[e.g. ][]{BossKeiser2010}. The presence of short-lived radionuclides in primitive meteorites \citep{OuelletteDeschHester2007,OuelletteDeschHester2009,DeschYoung2022} seems to support this scenario. These chemically processed elements, eventually inherited by the material within the protosolar disks, are now at the very base of life on Earth. Therefore, SNR-driven shocks may also play a role in replenishing molecular clouds with elements of pre-biotic interest.

Among these elements, Phosphorous (P) is of crucial biological importance, being it essential for the building up of cellular membranes, ribonucleic and deoxyribonucleic acids (RNA and DNA) and the Adenosine Triphosphate (ATP), i.e., the molecule that store energy for the human body \citep[see][for a review]{Fontani2024b}. Despite this essential role, P is highly depleted from the cold and dense gas of molecular clouds. The Solar elemental abundance of  Phosphorus is $\sim3\times10^{-7}$ relative to H \citep{Asplund2009}. In the gas phase in cold dense environments, earlier astrochemical models typically adopted depletion factors of $\sim1000$ \citep{Turner1990}, whereas more recent observational studies indicate that the phosphorus depletion in dense molecular gas may be larger than 60 \citep{rivilla2016,ziurys2018,Furuya2024}. As a consequence, the detection of P-bearing species in star forming regions is scarce. Toward these environments, only three P-bearing species have been detected so far: PN, PO and PO$^+$ \citep[e. g.][]{turnerBally1987,Ziurys1987,Tenenbaum2007,Rivilla2022} and all of them toward shocked and/or highly ionised environments \citep{rivilla2016,
Lefloch2016,Bergner2019,Wurmser2022,Rivilla2018,Rivilla2022,Scibelli2025}. 

\cite{Fontani2016}, \cite{Mininni2018} and \cite{Fontani2019} conducted a systematic survey of PN emission in a sample of both low- and high-mass star forming regions. The authors reported detection of sub-thermally excited (5-30 K) PN emission whose line profiles strongly correlate with that of the well-known shock tracer Silicon Monoxide (SiO) \citep{Fontani2019}. More recently, \cite{Fontani2024a} also highlighted a spatial association between PN and SiO emission toward the star forming region G31.41+00.31. The authors reported detection of PN toward the molecular outflows previously identified in the region \citep{beltran2006} but not toward the hot core. This suggests a tight correlation between PN and shocks and that hot gas phase chemistry may not be efficient in forming PN. Similar high-angular resolution studies performed toward high-mass \citep{Rivilla2020} and low-mass protostars \citep{Bergner2022} confirmed this results.

The first detection of PO$^+$ in molecular clouds has been reported by \cite{Rivilla2022} toward G$+$0.693, a dark cloud located nearby SgrB2 in the Galactic Center. More recently, \cite{Scibelli2025} have observed PO$^+$ toward a shocked low-mass core in the Perseus molecular cloud. No other detection of PO$^+$ emission have been reported so far in star forming regions. Strong SiO emission is observed toward both sources \citep{Lefloch1998,Zeng2020,Scibelli2025} and in both studies, the PO$^+$ is coupled with strong PN and PO emission. The abundance ratios between PN, PO and PO$^+$ have been proposed as possible diagnostics of energetic processes in the ISM, e. g. shocks, UV radiation and cosmic-ray ionisation \citep{jimenezserra2018}. In particular, the PO$^+$/PO and PO$^+$/PN ratios indicate that PO$^+$ is selectively enhanced in highly ionized gas, as its formation is dominated by ion–molecule reactions favoured by shocks and elevated cosmic-ray ionization rates \citep{Rivilla2022}.

While P is thought to be primarily produced in massive stars through neutron capture during the pre-supernova stage and subsequently ejected into the ISM during the supernova explosion \citep[e.g.][]{Woosley1995,Koo2013}, other nucleosynthetic processes, such as proton capture, have been proposed to contribute to its Galactic abundance \citep[e.g.][]{Caffau2011,Roederer2014,Maas2019,Nandakumar2022}. In this work, we conduct the first search of P-bearing species toward two well-known SNRs: IC443 and W44. Both sources are known to be interacting with molecular clouds and thus resemble the putative environment in which the Sun might have formed. IC443 is a core-collapse SNR, long believed to be $\sim$30 kyr old \citep[e. g. ][]{hewitt2006} but recently suggested to be significantly younger \citep[8 kyr;][]{ustamujic2021}. The SNR is located at a distance of 1.9 kpc \citep{ambrociocruz2017} and is interacting with a torus-shaped molecular cloud \citep{cornett1977,dickman1992}, suggested to be a translucent cloud \citep{vanDishoeck1998}. The strongest interaction site between the SNRs and the cloud is known as Clump G \citep{hewitt2006,troja2006,troja2008}. Toward this region, shock velocities up to 60-80 km s$^{-1}$ have been reported \citep{reach2019,Reach2024}.

W44 is also a core-collapse SNR, located at a distance of 2.9 kpc \citep{wootten1978} and estimated to be 20 kyr old \citep[W44;][]{wootten1977}. The SNR is known to be driving a slow shock of $\sim$10-20 km s$^{-1}$ \citep{sashida2013,cosentino2019} into a dense ridge at the edge of the infrared dark cloud G034.77-00.55 \citep{cosentino2018}. A relatively complex chemistry has been recently reported toward a region adjacent to the shock front, where both deuterated species and complex organic molecules have been detected \citep{cosentino2026}.

Toward both SNRs, strong SiO emission probing the shocked gas has been reported by \cite{cosentino2018,cosentino2019} and \cite{cosentino2022}, for W44 and IC443 respectively. Through high-angular resolution SiO(2-1) images obtained with the \textit{Atacama Large Millimeter/Submillimiter Array} (ALMA), the shock front driven by W44 has been spatially resolved and the shocked gas kinematics is compatible with that of a time-dependent CJ-type magneto-hydro-dynamic (MHD) shock \citep{cosentino2019}. Toward IC443, SOFIA and Spitzer far-Infrared (5-28 $\mu$m) observations of H$_2$ roto-vibrational and pure rotational lines have indicated the presence of multiple J and C type shocks toward Clump G \citep{reach2019,Reach2024,dellova2024}. The density enhancement caused by the two SNR-driven shocks is a factor $\geq$10, with final post-shocked gas density $>$10$^5$cm$^{-2}$ \citep{cosentino2019,cosentino2022}. Recent ALMA observations of the dense gas tracer N$_2$H$^+$(1-0) have revealed the presence of a dense core population at the starless or pre-stellar stage whose formation has been triggered by the W44-driven shock  \citep{cosentino2025a}. Toward IC443, indirect evidence of the presence of cores has been suggested, but the shock has been speculated to be hindering star formation in these objects \citep{Reach2024}.

Toward these two SNRs, we here investigate the presence of PN and PO$^+$ emission and their correlation with the shocked gas emission. The paper is organised as follows. In Section~\ref{sec:ObsAndData} we describe the observational strategy. In Section~\ref{subsec:results}, we discuss the method used to detect the species of interest and derive molecular abundances. In Section~\ref{sec:models}, ~\ref{sec:discussion} and ~\ref{sec:conclusions}, we discuss our result and present our conclusions.

\section{Observations and data}\label{sec:ObsAndData}
\begin{figure*}[!htpb]
    \centering
    \includegraphics[width=1\linewidth]{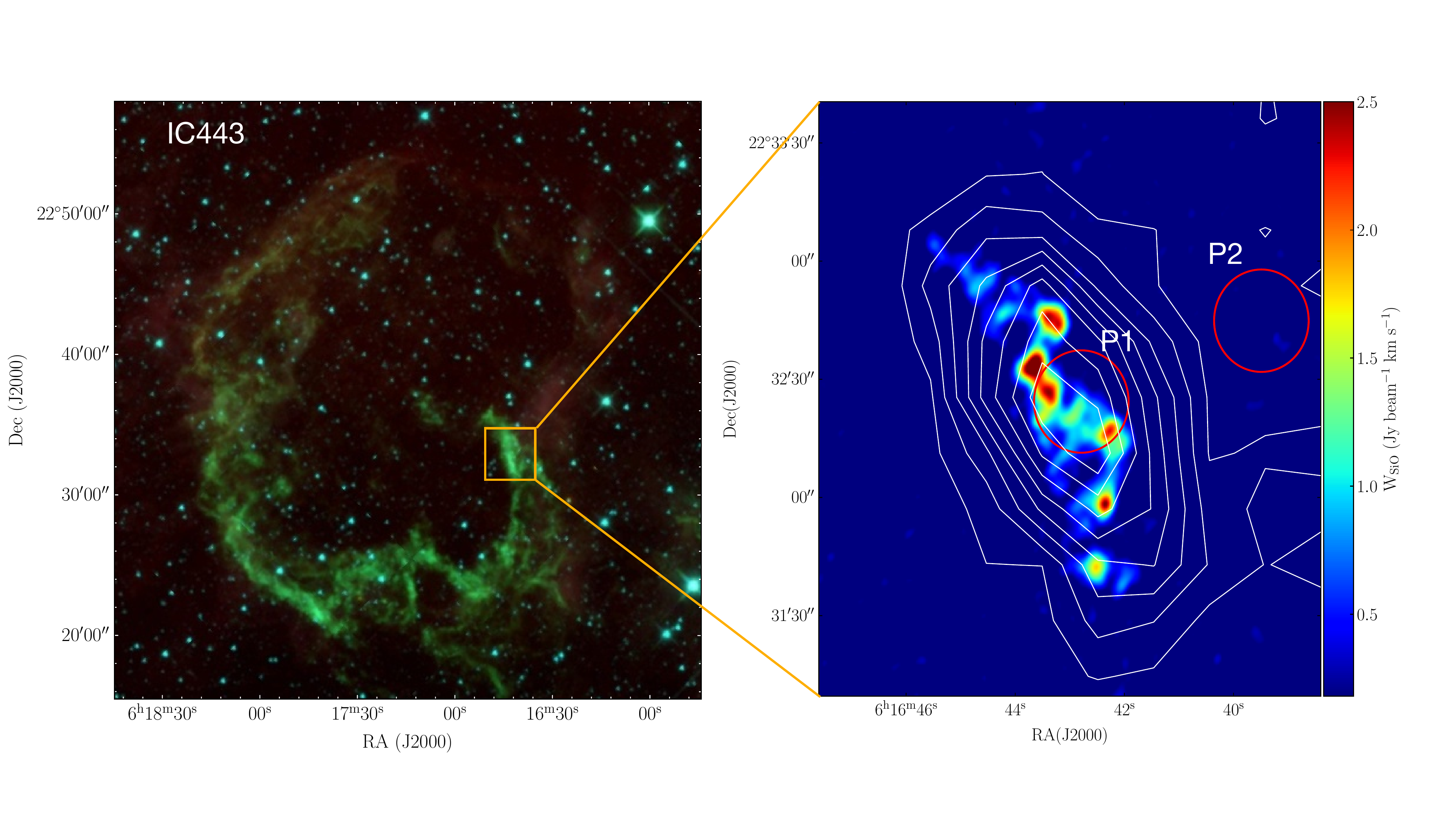}
    \includegraphics[width=1\linewidth]{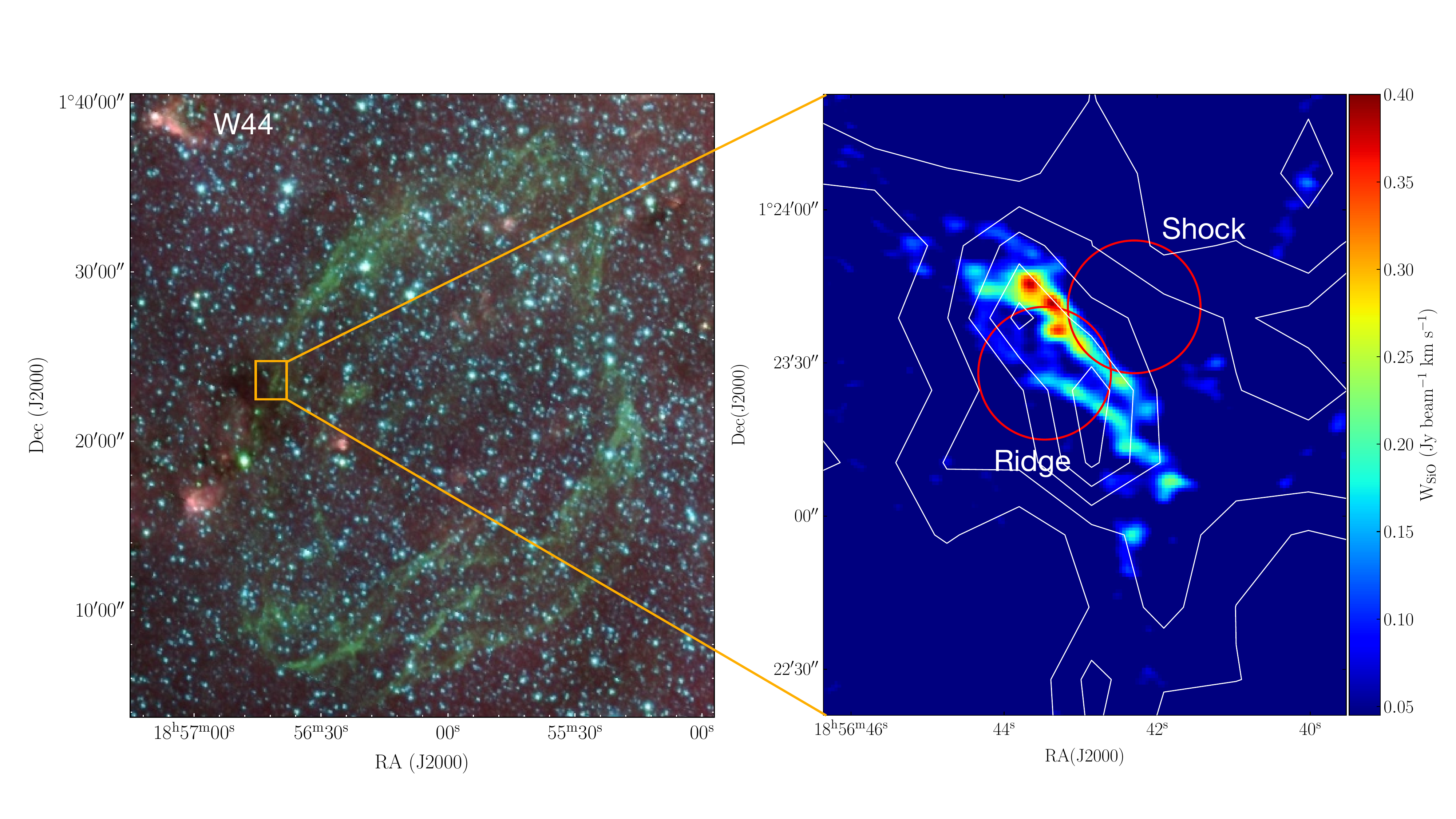}
    \caption{Three-colour image (top left; red= 22 $\mu$m; green = 4.6 $\mu$m; blue = 3.4 $\mu$m) of the SNR IC443 \citep[WISE all-sky survey][]{wright2010} and the SNR W44 (bottom left; red= 8 $\mu$m; green = 5.3 $\mu$m; blue = 4.5 $\mu$m) from the Spitzer GLIMPSE survey \citep{churchwell2009}. On the right panels, zoom-in views of the regions of interest are shown. The integrated intensity above the 3$\sigma$ level of the SiO(2-1) emission seen with ALMA at a resolution of 3$^{\prime\prime}$ toward IC443 (top right; $\sigma$=0.06 Jy beam$^{-1}$ km s$^{-1}$; integration range is -20,20 km s$^{-1}$) and W44 (bottom right; $\sigma$=0.02 Jy beam$^{-1}$ km s$^{-1}$; integration range is 38,47 km s$^{-1}$) is shown in color scale. In both panels, the white contours correspond to the SiO(2-1) emission obtained with the IRAM-30m \citep{cosentino2018}. The red circle highlights the positions analysed in this work and for which high-sensitivity 3 mm spectra have been obtained. The circle aperture corresponds to 26$^{\prime\prime}$, i.e. the beam aperture at the PO$^+$ frequency.}
    \label{fig:fig1}
\end{figure*}

In March 2020 and 2021, we obtained 3 mm high-sensitivity single pointing observations towards the shocked gas around the SNRs W44 and IC443. Toward each source, two positions were selected: one corresponding to the SiO emission peaks reported by \cite{cosentino2018} and \cite{cosentino2022}, and one additional  region toward each SNR, offset with respect to the SiO peak. Toward W44, the pointing here investigated corresponds to the region named Ridge and Shock in \cite{cosentino2023,cosentino2025a}. Toward IC443, both the on- and off-shock positions, named IC443-P1 and IC443-P2 respectively, are located toward clump G. Observations were performed using the EMIR E90 receiver on the 30m antenna at the Instituto de Radioastronom\'ia Millim\'etrica (Pico Veleta, Spain), in position switching mode. By using tuning frequencies of 74.0 \textbf{GHz} and 81.4 GHz, we obtained a continuous spectral coverage from 71.7 GHz to 102.6 GHz. The FTS spectrometer was used with 200 kHz channel width mode, corresponding to velocity resolution 0.6-0.8 km s$^{-1}$ and beam aperture in the range 24-34$^{\prime\prime}$. At the kinematic distances of W44 (2.9 kpc) and IC443 (1.9 kpc) these apertures correspond to physical scales in the range 0.3-0.5 \textbf{pc} and 0.2-0.3 \textbf{pc}, respectively.

Intensities were measured in units of antenna temperature and converted into main-beam brightness temperature, T$_{\rm{mb}}$, using beam and forward efficiencies of 0.81 and 0.95\footnote{https://publicwiki.iram.es/Iram30mEfficiencies}. The final spectra were produced using the {\sc class} software within the {\sc gildas} package\footnote{See http://www.iram.fr/IRAMFR/GILDAS} by smoothing to a velocity resolution of 1.2-1.6 km s$^{-1}$. The final achieved rms per channel is in the range 5-3 mK for W44 and 4-2 mK for IC443. This information, together with the pointing positions and off positions are reported in Table~\ref{tab:tab1}. In this paper, we investigate the presence of the J=2-1 PN and PO$^+$ emission lines at 93.98 GHz and 94.05 GHz respectively. At these frequency, the IRAM-30m beam size is 26$^{\prime\prime}$, the velocity resolution after smoothing is 1.2 km s$^{-1}$ and the achieved rms per channel is 2 mK toward IC443 and 5 mK toward W44. We note that no PO transition lines fall within our frequency broad bandwidth. We will present additional species of pre-biotic interest (S-bearing, N-bearing, O-bearing etc.) in a series of forthcoming papers. 

In Figure~\ref{fig:fig1}, the morphologies of the two SNRs are shown by means of three-color Infrared images from the WISE all-sky survey \citep{wright2010} and the Spitzer GLIMPSE survey \citep{churchwell2009}, for IC443 and W44 respectively. On the rightmost panels, zoom-in views of the ongoing shock interactions are shown by means of the SiO(2-1) images observed by ALMA, and with overlaid the on-shock and off-shock positions analysed in this work (red circles). For both maps, the achieved angular resolution is $\sim$3$^{\prime\prime}$, the spectral resolution is 0.2 km s$^{-1}$ and the achieved sensitivity is 10 mJy beam$^{-1}$. The ALMA SiO image toward W44 (PI: I. Jim\'enez-Serra; project code: 2016.1.01363.S) have been obtained by combining ALMA and ACA observations. For a detailed discussion of the observing setup we refer to \cite{cosentino2019}, where the observations have been first presented. The ALMA SiO image toward IC443 (PI: G. Cosentino; project code: 2016.1.01363.S) have been obtained with a similar strategy to that reported by \cite{cosentino2019} and by combining ALMA, ACA and Total Power data. The data will be fully presented in a paper in preparation focused on the shocked gas kinematics seen with ALMA (Cosentino et al. in prep.). In the right panels, we also overlay the SiO(2-1) integrated intensity emission contours obtained with the IRAM-30m using the observing strategy described in \cite{cosentino2018}. 

\begin{table*}
    \centering
    \caption{List of observed positions, covered frequency range and beam sizes, achieved spectral resolution and sensitivities for the IRAM-30m observations.}
    \begin{tabular}{lcccccccc}
    \hline
    \hline
        Source & \multicolumn{2}{c}{Position} & \multicolumn{2}{c}{Off-Position} & Frequency & $\delta$v & rms & $\theta$\\
        &  RA(J2000) & Dec(J2000) &  RA(J2000) & Dec(J2000)\\
        & (hh:mm:ss)  & (dd:mm:ss) & (hh:mm:ss)  & (dd:mm:ss) & (GHz) & (km s$^{-1}$) & (mK) & (${\prime\prime}$)\\
    \hline
    IC443-P1     &06:16:42.8  &22:32:24.4 & 06:16:15.8 & 22:32:22.4 &71.7-102.6 &1.2-1.6 &4-2 &34-24\\
    IC443-P2     &06:16:39.5  &22:32:44.8 & 06:16:15.8 & 22:32:22.4 &71.7-102.6 &1.2-1.6 &4-2 &34-24\\
    W44-Shock    &18:56:42.3  &01:23:40.0 & 18:57:01   & 01:22:25.0 &71.7-102.6 &1.2-1.6 &5-3 &34-24\\
    W44-Ridge    &18:56:43.4  &01:23:28.0 & 18:57:01   & 01:22:25.0 &71.7-102.6 &1.2-1.6 &5-3 &34-24\\
    \hline
    \end{tabular}
    \label{tab:tab1}
\end{table*}

\section{PO$^+$ and PN emission in IC443 and W44}\label{subsec:results}
\subsection{Detection and molecular abundances}\label{subsec:abund}
Across both SNRs, we have used the tool \textit{Spectral Line Identification and Modelling}, {\sc slim} within the {\sc madcuba} software\footnote{Madrid Data Cube Analysis on ImageJ is a software developed at the Center of Astrobiology in Madrid; http://cab.intacsic.es/madcuba/Portada.html} \citep{martin2019}, to identify the species of interest in the broadband spectra and to obtain line models of the detected transitions in Local-Thermodynamic-Equilibrium (LTE) conditions. The best LTE models obtained with {\sc slim} allow to estimate the species line profile parameters such as centroid velocity, v$_0$, line full width half maximum (FWHM), $\Delta$v, and peak intensity, T$_{\rm{peak}}$. With this method, already described in \cite{cosentino2026}, we detect the PO$^+$(2-1) transition at 94.047798 GHz toward the two position named W44-Shock and IC443-P1. The transition is known to be isolated (not blended) from other species transitions \citep{Rivilla2022} and we have verified this in {\sc slim}.

\begin{table}[!htpb]
    \centering
    \caption{Molecular species, transition, frequency, E$_{\mathrm{up}}$, g$_{\mathrm{u}}$, A$_{\mathrm{ul}}$ of all the transitions of interest.}
    \begin{tabular}{llcccc}
    \hline
    \hline
        Species & Transition & $\nu$ & E$_{\mathrm{low}}$ & g$_{\mathrm{u}}$ & A$_{\mathrm{ul}}$ \\
                             & & (GHz) & (K) & & (s$^{-1}$)\\
    \hline
        PO$^+$ & J=2-1 & 94.0478 & 1.569 & 5 & 4.5$\times$10$^{-5}$\\
        PN     & J=2–1, F=2–2 & 93.9782 &1.569 & 5 & 7.3$\times$10$^{-6}$ \\
               & J=2–1, F=1–0 & 93.9785 &1.569 & 3 & 1.6$\times$10$^{-5}$\\
               & J=2–1, F=2–1 & 93.9798 &1.569 & 5 & 2.2$\times$10$^{-5}$\\
               & J=2–1, F=3–2 & 93.9799 &1.569 & 7 & 2.9$\times$10$^{-5}$\\
               & J=2–1, F=1–2 & 93.9808 &1.569 & 3 & 8.1$\times$10$^{-7}$\\
               & J=2–1, F=1–1 & 93.9823 &1.569 & 3 & 1.2$\times$10$^{-5}$\\
        SiO    & J=2-1        & 86.8469 &1.448 & 5 & 2.9$\times$10$^{-5}$\\
    \hline
    \end{tabular}
    \label{tab:spectroscopic}
\end{table}

\begin{figure*}[!htpb]
    \centering
    \includegraphics[width=0.5\linewidth]{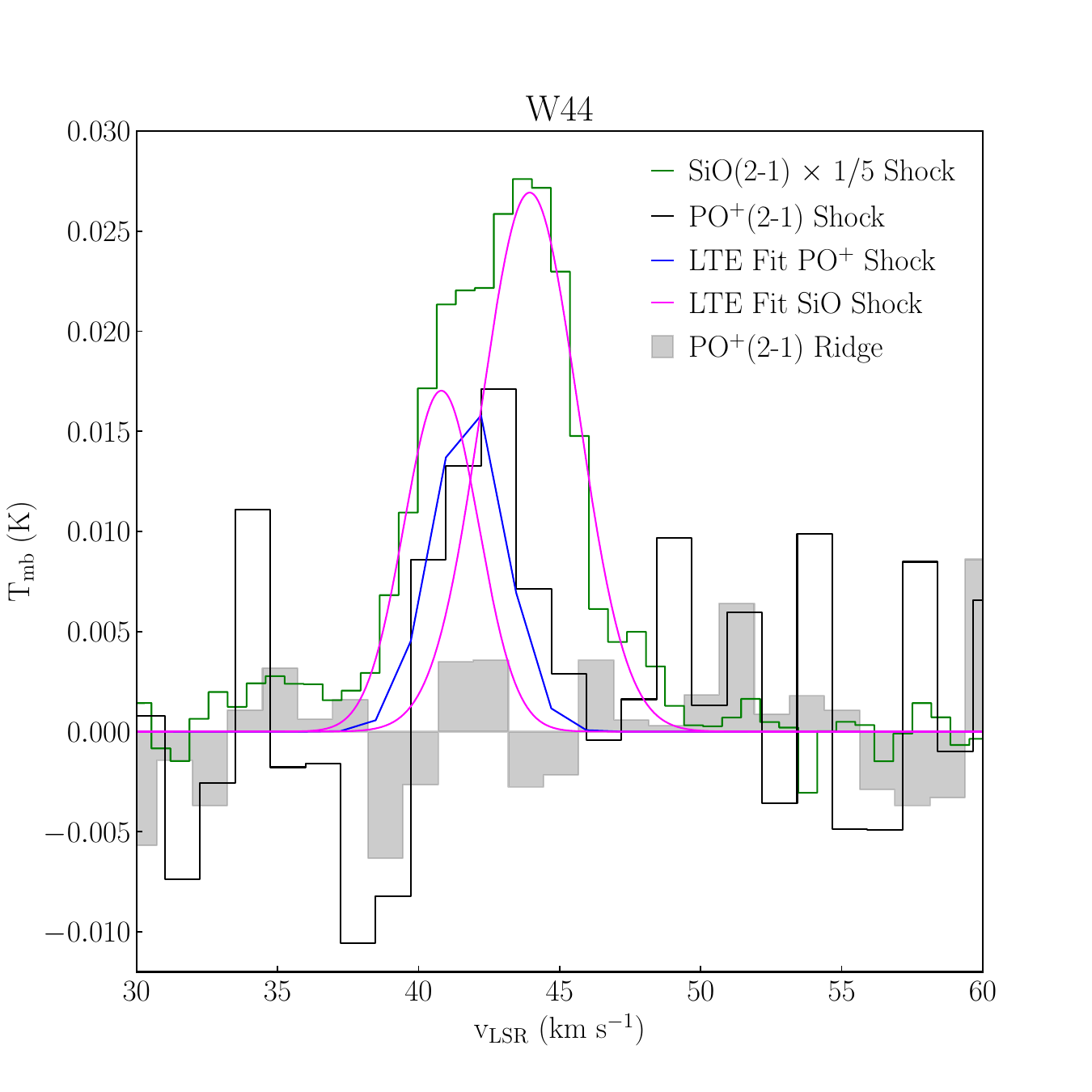}\includegraphics[width=0.5\linewidth]{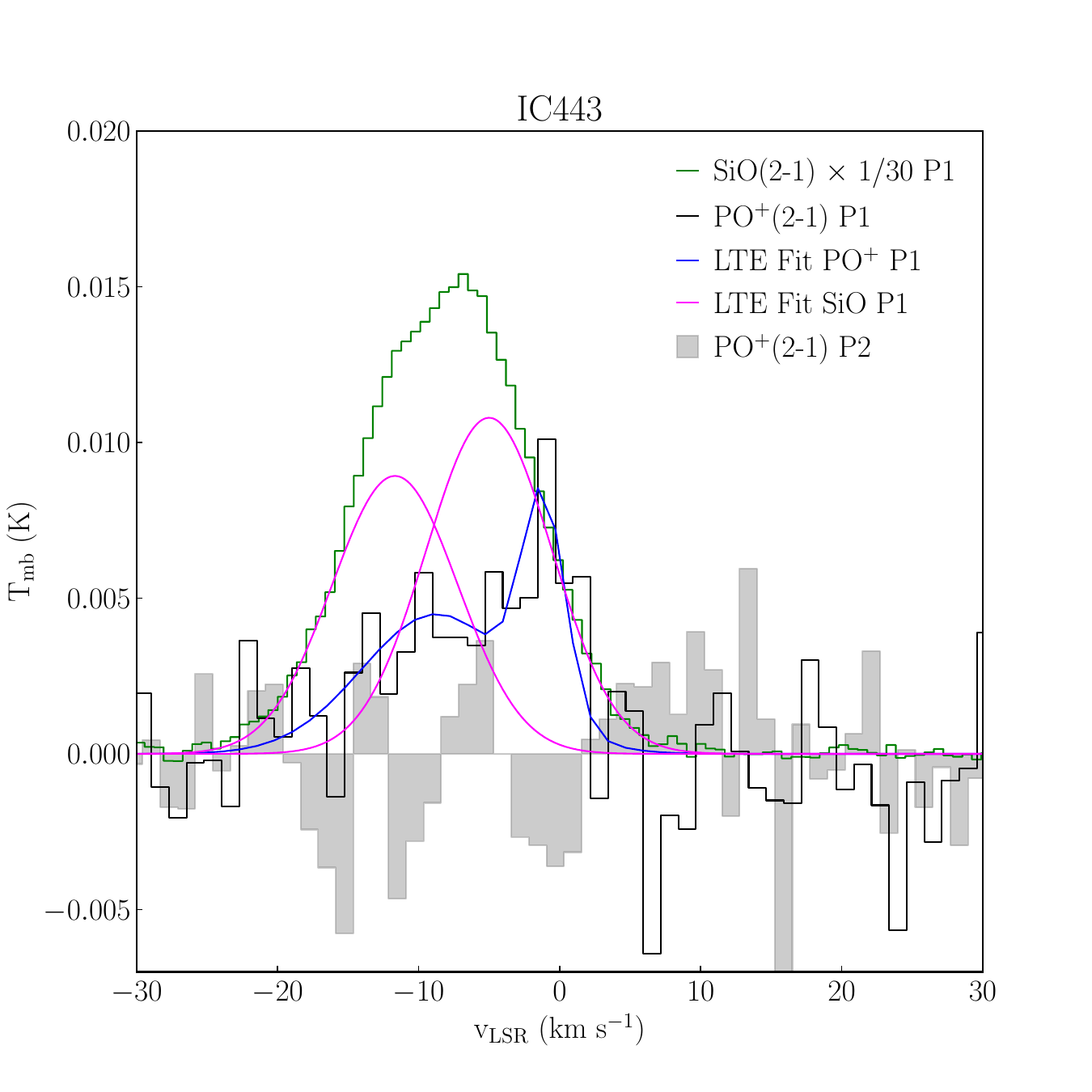}
    \includegraphics[width=0.5\linewidth]{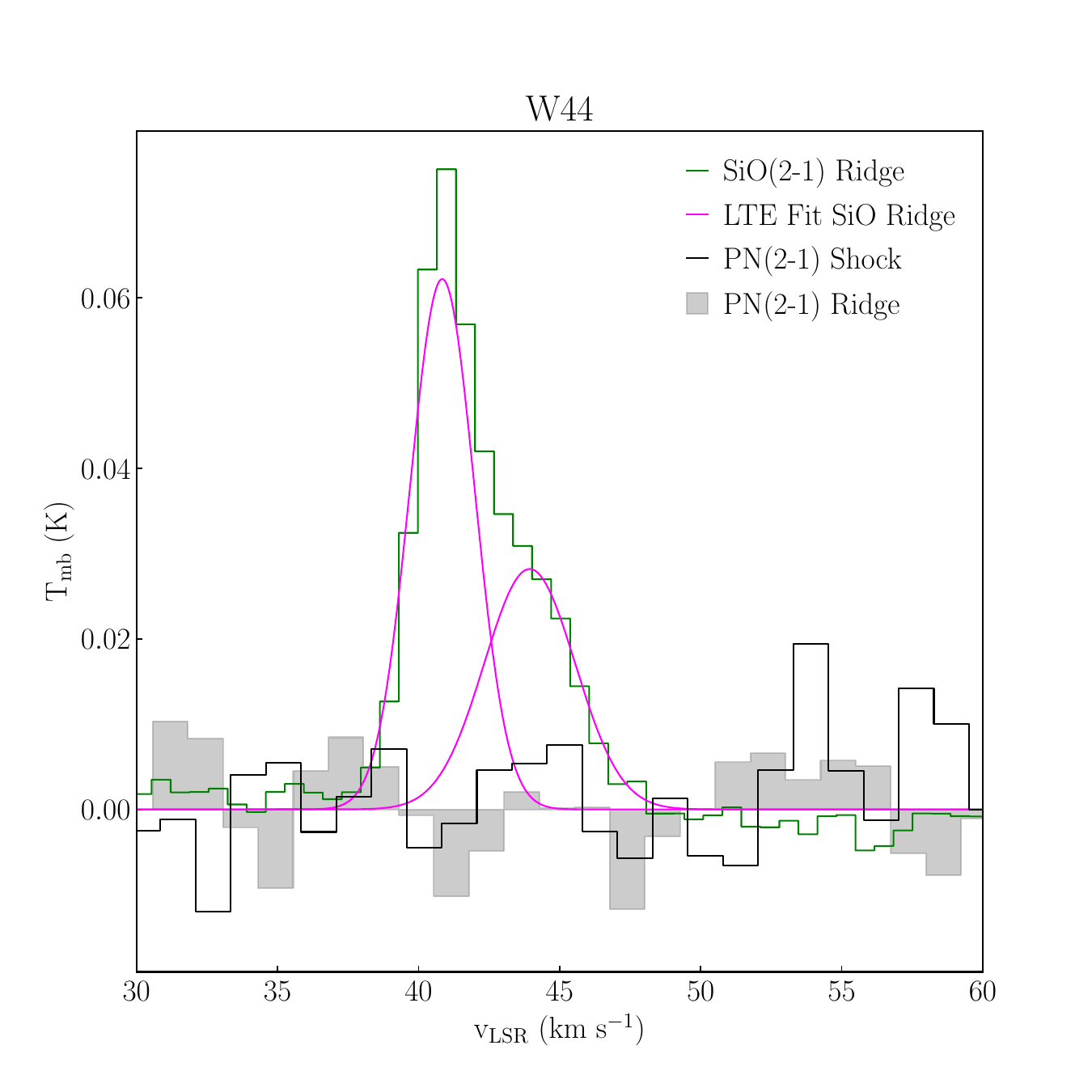}\includegraphics[width=0.5\linewidth]{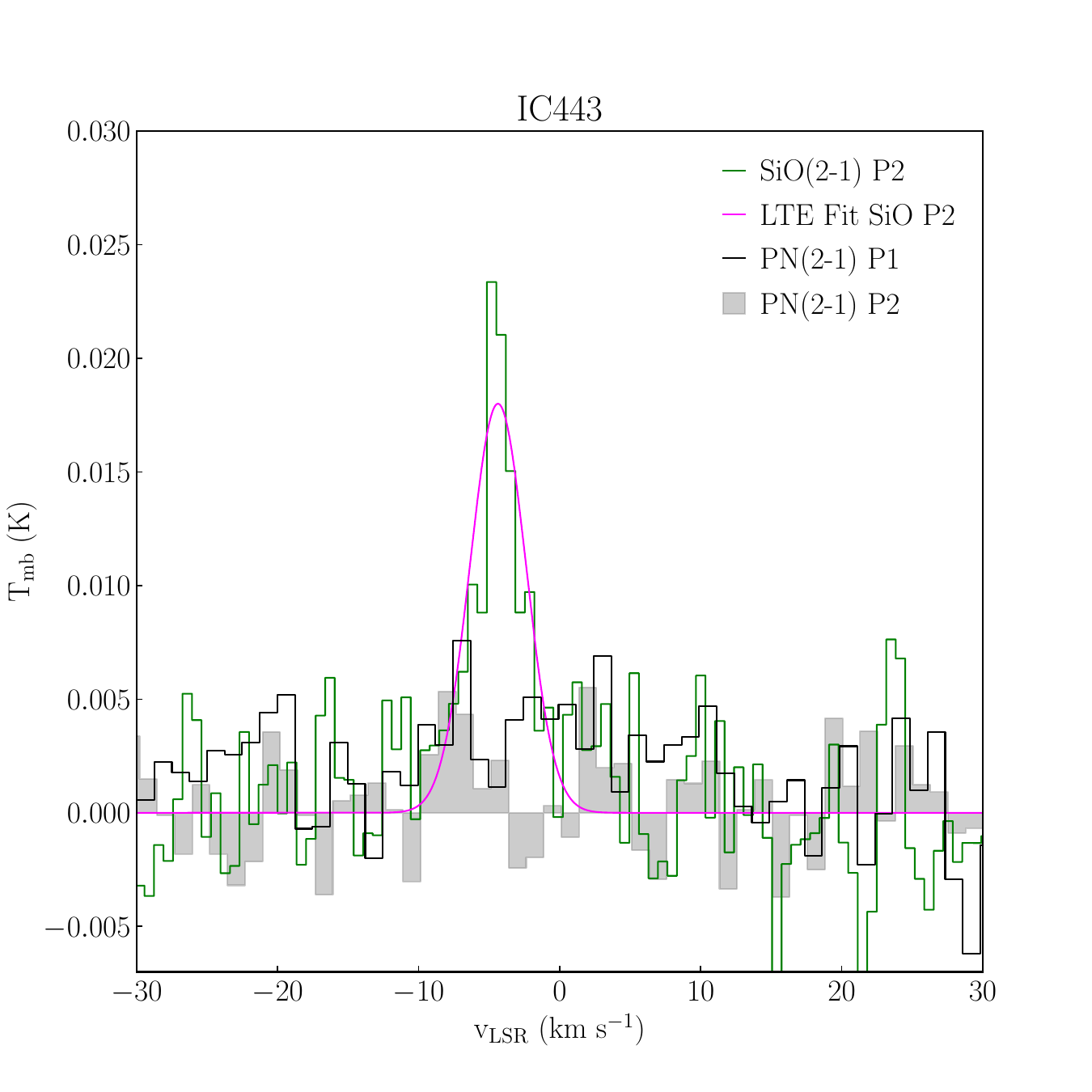}
    \caption{\textit{Top:} Spectra of the J=2-1 PO$^+$ transition obtained toward W44 (left panels) and IC443 (right panels). In all panels, the black lines show the PO$^+$ spectra obtained toward W44-Shock (left) and IC443-P1 (right). The blue curves show the best LTE model obtained with {\sc slim} in {\sc madcuba} toward these two positions. The green curves indicate the SiO(2-1) spectra obtained toward the two positions. For a better comparison, the SiO spectra have been scaled by a factor of 1/5 for W44 and 1/30 for IC443. Finally, the magenta lines show the SiO Gaussian fitting obtained toward the two positions. The gray shadows show the PO$^+$ spectra toward W44-Ridge and IC443-P2. \textit{Bottom:} Same as the top panels but with black lines and grey curves showing the PN spectra toward both W44 (left) and IC443 (right).}
    \label{fig:spectra}
\end{figure*}

At the spectral resolution of 1.2 km s$^{-1}$, the PO$^+$ emission toward W44-Shock is best reproduced by a single velocity component at $\sim$42 km s$^{-1}$. The line intensity peak is T$_{\rm{peak}}\sim$17 mK, hence we infer a signal-to-noise ratio (S/N) of $\sim$3.5, calculated as the ratio between T$_{\rm{peak}}$ and the rms (5 mK). We also estimate the integrated S/N ratio to be $\sim$5.7, as the ratio between the line integrated intensity and the integrated noise, A$_{\mathrm{rms}}$\footnote{ A$_{\mathrm{rms}}$ = rms $\times \delta$v $\times\sqrt{N_{ch}}$, where $\delta$v is the spectrum velocity resolution (1.2 km s$^{-1}$) and $N_{ch}$ is the number of channels within the line width, $\Delta$v}.
Toward IC443-P1, the PO$^+$ emission is best reproduced by two velocity components: a broader component at -8.4 km s$^{-1}$ and a narrower, slightly brighter component at -1.3 km s$^{-1}$. From the line profile parameters obtained with {\sc slim}, the intensity and integrated S/N are 2.5 and 5 for the broad component at -8.4 km s$^{-1}$ and 3.5 and 5.5 for the narrower component at -1.3 km s$^{-1}$. We therefore conclude that the PO$^+$ emission is detected with a 3$\sigma$ level confidence toward both SNRs. Toward the positions named W44-Ridge and IC443-P2 no significant PO$^+$ emission is detected. In Figure~\ref{fig:spectra}, the PO$^+$ spectra across the four positions are shown along with the SiO(2-1) IRAM-30m spectra obtained toward W44-Shock and IC443-P1, but with the native velocity resolution of 0.6 km s$^{-1}$.

Using the spectroscopic parameters available in the Jet Propulsion Laboratory catalogue\footnote{https://spec.jpl.nasa.gov/} \citep[JPL;][]{Petrmichl1991,Pickett1998}, and reported in Table~\ref{tab:spectroscopic}, the LTE model fitting done in {\sc slim} can also estimate the PO$^+$ column density, N(PO$^+$) and the excitation temperature, T$_{\mathrm{ex}}$. In our case, since only one transition is available, T$_{\mathrm{ex}}$ needs to be fixed. Hence, we assume excitation temperature T$_{\rm{ex}}$= 10 K, consistent with that inferred by \cite{cosentino2022} and \cite{cosentino2018} through a multi-transition analysis of the SiO emission in IC443, and CH$_3$OH emission in W44 respectively. The assumed excitation temperature is also consistent with that inferred by \cite{dellova2020} through population diagrams of CO emission in IC443. In our calculation, we also account for the uncertainties in fixing T$_{\rm ex}$. For this, we estimate the PO$^+$ column densities using $T_{\rm ex}$ in the ranges 5--15 K, consistent with the $T_{\rm ex}$ uncertainty ranges reported by \cite{cosentino2022} and \cite{cosentino2018} for the two SNRs. Hence, we considered the maximum variation with respect to the value obtained at $T_{\rm ex}=10$ K as the systematic uncertainty associated with the excitation temperature. This corresponds to an uncertainty of approximately 30\%. We then sum this contribution in quadrature with the statistical uncertainty from the spectral fitting. The resulting total uncertainties are reported in Table~\ref{tab:tab2}.\\

We note that a filling factor of 1 was used for both sources, since the PO$^+$ spatial distribution is unknown. From the ALMA SiO images shown in Figure~\ref{fig:fig1}, and assuming a tight correlation between P-bearing species and SiO morphology \citep{Fontani2019}, a filling factor $<$1 may be more accurate. Hence, our PO$^+$ column density estimates may be regarded as lower limits. Higher-angular resolution observations of the PO$^+$ emission are needed to constrain the true size of the emitting regions. 

We estimate column densities N(PO$^+$)=9$\times$10$^{10}$ cm$^{-2}$ in W44-Shock and 11$\times$10$^{10}$ cm$^{-2}$ in IC443-P1, obtained by summing the contribution from the two velocity components. In order to estimate the PO$^+$ fractional abundance toward W44, we consider molecular Hydrogen column density, N(H$_2$), of (1.1$\pm$0.3)$\times$10$^{22}$ cm$^{-2}$ toward W-44 Shock and (1.4$\pm$0.4)$\times$10$^{22}$ cm$^{-2}$ toward W44-Ridge, as estimated by the mass surface density values reported by \cite{cosentino2023}. In IC443, \cite{dellova2020} used multiple CO transitions and a conversion factor [H$_2$]/[CO]=10$^4$ to infer the N(H$_2$) value toward the region here analysed. The authors estimate N(H$_2$) for the bulk, blue-shifted and red-shifted CO emission toward this area. We consider the sum of the three contributions and adopt a value of N(H$_2$)=(2.6$\pm$0.7)$\times$10$^{21}$ cm$^{-2}$, and we calculated the uncertainty by following the standard error propagation rules. We note that this value is also consistent with that measured by \cite{dellova2024} through rotational diagrams of H$_2$ pure rotational transitions. Following this method, the estimated PO$^+$ abundances are $\chi$(PO$^+$)=(8$\pm$3)$\times$10$^{-12}$ toward W44-Shock and $\chi$(PO$^+$)=(5$\pm$2)$\times$10$^{-11}$ toward IC443-P1. Finally, using the 3$\times$A$_{\mathrm{rms}}$ values and assuming linewidth similar to that observed toward the other positions, we also estimate the PO$^+$ upper limits toward W44-Ridge and IC443-P2 to be $\chi\leq$2$\times$10$^{-12}$ and $\chi\leq$1$\times$10$^{-11}$ respectively.  Unfortunately,  N(H$_2$) estimates toward IC443-P2 position are not available in literature, hence we have adopted the same N(H$_2$) value for both IC443-P1 and IC443-P2. All the obtained quantities are reported in Table~\ref{tab:tab2} for both W44 and IC443, along with the obtained SiO abundances. Similarly to PO$^+$, the SiO line profile parameters and column densities have been obtained by fitting the SiO lines in {\sc slim}. Toward all positions but IC443-P2, the SiO emission is best reproduced by a two velocity components fitting. The final SiO abundances have been obtained by summing the contributions from the two velocity components.

Toward the four positions analysed in this work, we do not detect any PN emission above the 3$\sigma$ level. The portion of the spectra with the putative PN emission are shown in Figure~\ref{fig:spectra}. Hence, following the method described above for the PO$^+$ upper limits, we estimate the PN abundance upper limits toward the four positions to be $\leq$3$\times$10$^{-12}$ cm$^{-2}$ toward W44 and $\leq$10$^{-11}$ cm$^{-2}$ toward IC443. All the estimated values of column densities and abundances are listed in Table~\ref{tab:tab2}.\\

Finally, to assess the impact of possible sub-thermal excitation, we also performed non-LTE radiative-transfer calculations with RADEX \citep{tak2007}, using the collisional coefficients available in the LAMDA database for PO$^+$ \citep{Tonolo2024} and PN \citep{tobola2007}, respectively (Appendix~\ref{app:radex}). Since only one rotational transition is covered for both species, the excitation conditions cannot be simultaneously constrained. Nevertheless, for densities representative of the compressed gas \citep[n(H$_2$)>10$^4$ cm$^{-3}$][]{cosentino2019,cosentino2022} and kinetic temperatures up to $\sim$30 K (see Figure~\ref{fig:radex_po}), the PO$^+$ column densities inferred from the non-LTE calculations are comparable, within a factor of a few, to the LTE estimates derived above. We therefore conclude that possible departures from LTE may change the PO$^+$ column densities derived in this work by a factor of few.

\begin{table*}[]
    \centering
    \caption{Centroid velocity, linewidth, peak intensity, total column density and molecular abundance with respect to H$_2$ obtained for all the species of interest. Molecular abundances have been calculated by assuming N(H$_2$)= (11$\pm$3)$\times$10$^{21}$ cm$^{-2}$ toward W44-Shock, N(H$_2$)= (14$\pm$4)$\times$10$^{21}$ cm$^{-2}$ toward W44-Ridge and N(H$_2$)= (2.6$\pm$0.7)$\times$10$^{21}$ cm$^{-2}$ toward IC443-P1 and IC443-P2.}
    \setlength{\tabcolsep}{4pt}
    \begin{tabular}{lccccc|ccccc|c}
    \hline
    \hline
    Position &v$_0$ & $\Delta$v & T$_{\rm{peak}}$ & N(PO$^+$) & $\chi$(PO$^+$) & v$_0$ & $\Delta$v & T$_{\rm{peak}}$ & N(SiO) & $\chi$(SiO) & $\chi$(PN)\\
             & (km s$^{-1}$) & (km s$^{-1}$) & (mK) & (10$^{10}$ cm$^{-2}$) & (10$^{-11}$) & (km s$^{-1}$) & (km s$^{-1}$) & (mK) & (10$^{12}$ cm$^{-2}$) & (10$^{-9}$) & (10$^{-11}$)\\
    \hline
     W44-Shock &41.7$\pm$0.5 &3$\pm$1   &17$\pm$5  &9$\pm$4   &0.8$\pm$0.4 &43.6$\pm$0.7 &4$\pm$1& 129$\pm$5 &1.2$\pm$0.5 &0.17$\pm$0.05 &$<$0.3\\
               &$\cdots$ &$\cdots$   &$\cdots$  &$\cdots$   &$\cdots$  &40.3$\pm$0.9     &3$\pm$1  &75$\pm$5 & 0.6$\pm$0.3 &0.17$\pm$0.05 &$<$0.3\\ 
     W44-Ridge &$\cdots$     &$\cdots$  &$\cdots$  &$\cdots$  &$\leq$0.3   &43.6$\pm$0.4 &3.8$\pm$0.7 &141$\pm$5 &1.2$\pm$0.3 &0.23$\pm$0.07 &$<$0.3\\
               &$\cdots$     &$\cdots$  &$\cdots$  &$\cdots$  &$\leq$0.3   &40.5$\pm$0.4 &2.7$\pm$0.5 &310$\pm$5 &1.9$\pm$0.3 &$\cdots$ &$\cdots$\\
     IC443-P1  &-8.4$\pm$2.5 &11$\pm$5  &4$\pm$2   &8$\pm$3   &5$\pm$2     &-13$\pm$1 &10$\pm$1  &220$\pm$2 &5$\pm$2 &6$\pm$1 &$<$1\\
               &-1.3$\pm$0.6 &4.0$\pm$1 &7$\pm$2   &4$\pm$2   &$\cdots$    &-6$\pm$1  &11$\pm$1 &370$\pm$2  &10$\pm$2        &$\cdots$\\
     IC443-P2  &$\cdots$     &$\cdots$  &$\cdots$  &$\cdots$  &$\leq$1     &-4.7$\pm$0.3 &4.7$\pm$0.6 &18$\pm$3 &0.20$\pm$0.02 &0.07$\pm$0.02 &$<$1\\
    \hline
    \end{tabular}
    \label{tab:tab2}
\end{table*}

\subsection{Relation between SiO and PO$^+$}
We now investigate possible similarities between the SiO and PO$^+$ line profiles as well as possible correlation between the two species molecular abundances. From Figure~\ref{fig:fig1}, the SiO spectrum obtained toward W44-Shock show SiO emission peaks at 40 km s$^{-1}$ and $\sim$44 km s$^{-1}$. These velocities correspond respectively to the shocked gas associated respectively with the J- and C-type part of the 20 km s$^{-1}$ shock driven by the SNR \citep{cosentino2019}. The $\sim$42 km s$^{-1}$ PO$^+$ centroid velocity here reported is consistent with the lowest velocity of the gas shocked by the J-type shock i.e., a region of high-compression. The FWHM of both the PO$^+$ and SiO emission is $\sim$3 km s$^{-1}$.\\

Toward IC443-P1, the SiO spectrum also presents two velocity components. Both of them appear systematically red-shifted with respect to the PO$^+$ velocity components. The presence of multiple overlapping C and CJ-type shocks in this region \citep{reach2019,Reach2024} may explain the presence of multiple PO$^+$ velocity components, which may correspond to the most compressed shock layers of two different shocks. We note that the narrower PO$^+$ velocity component at -1.3 km s$^{-1}$ is indeed consistent with the most powerful region of the shock, as reported by \cite{cosentino2022}.\\ 

\begin{figure}
    \centering
    \includegraphics[width=\linewidth]{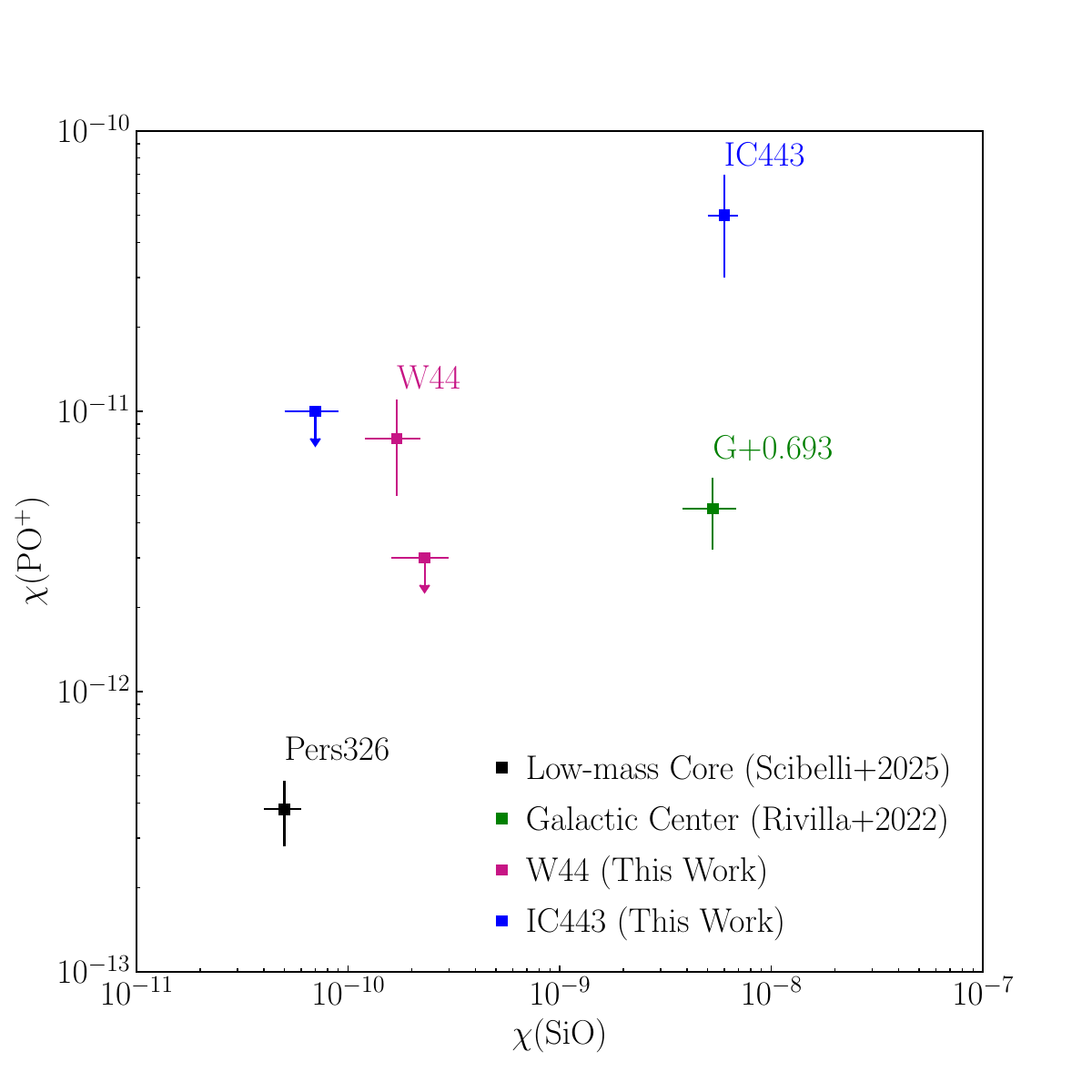}
    \caption{PO$^+$ abundance, $\chi$(PO$^+$), as a function of the SiO abundance, $\chi$(SiO) obtained toward the two SNRs W44 (magenta) and IC443 (blue). The upper limits estimated toward the Ridge and P2 are also indicated. For comparison, the estimates reported by \cite{Rivilla2022} toward the molecular cloud G$+$0.639 (green) in the Galactic Center and by \cite{Scibelli2025} toward the low-mass shocked core (black) in the Perseus molecular cloud are also reported.}
    \label{fig:abund}
\end{figure}

In Figure~\ref{fig:abund}, we also compare the $\chi$(SiO) and $\chi$(PO$^+$) abundances obtained toward the four regions here analysed and the two values reported in the literature \citep{Rivilla2022,Scibelli2025}. The SiO molecular abundance toward Pers326 has been obtained by considering the SiO integrated intensity reported by \cite{Scibelli2025}. The $\chi$(SiO) toward G$+$0.693 has been obtained from \cite{Massalkhi2023}. The few detections of PO$^+$ available are not sufficient to draw statistically significant conclusions on the trend between $\chi$(SiO) and $\chi$(PO$^+$). However, there may be a hint of positive correlation between the two quantities. Notably, the $\chi$(PO$^+$) measured toward W44 are consistent with those reported by \cite{Rivilla2022} toward the Galactic Center molecular cloud G$+$0.693, but the $\chi$(SiO) estimated toward W44 are $\sim$2 orders of magnitude lower. On the other hand, the $\chi$(SiO) toward IC443 is similar to that reported by G$+$0.693, but the PO$^+$ abundance is one order of magnitude larger.

\subsection{Relation between PO$^+$, PN and SiO}
\begin{figure}
    \centering
    \includegraphics[width=\linewidth, trim = 2cm 2cm 1cm 10cm, clip=True]{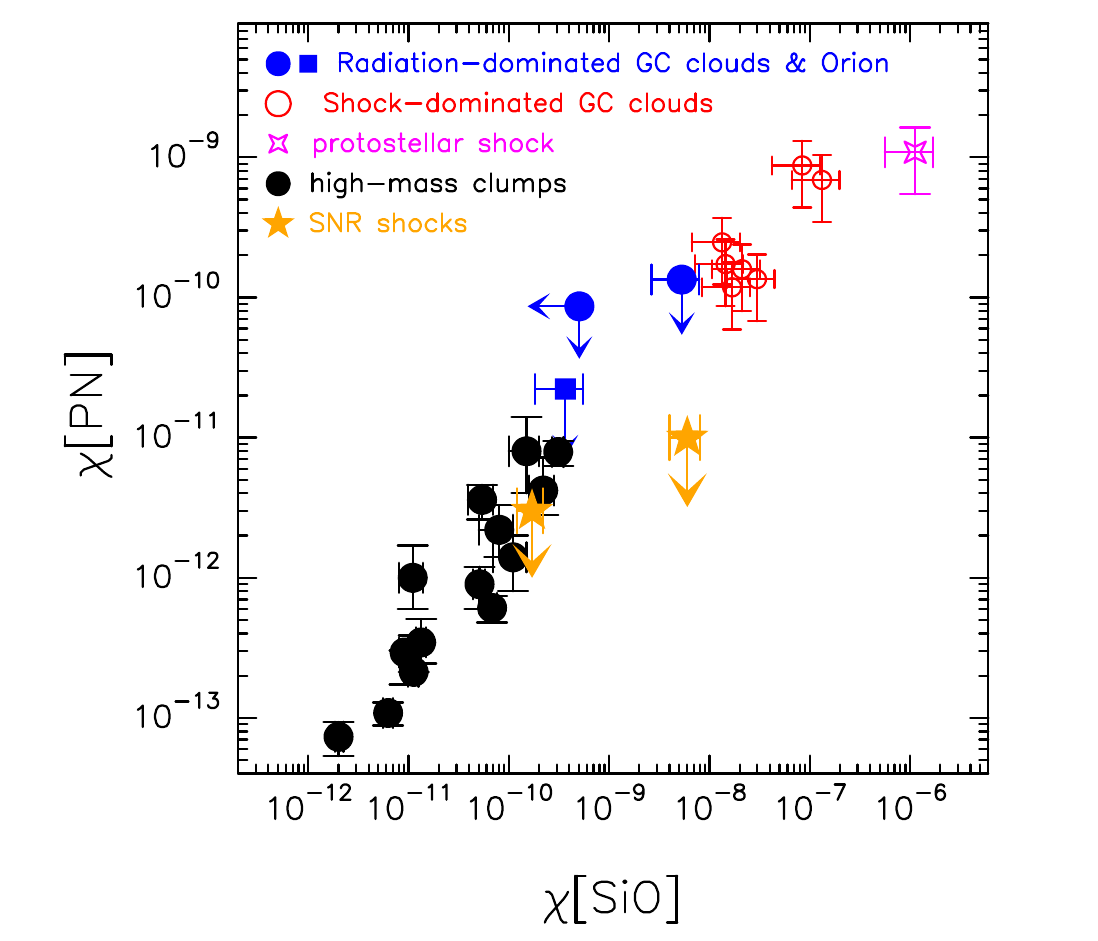}
    \caption{Fractional abundance of PN, $\chi$(PN), against that of SiO, $\chi$(SiO), in different Galactic star formation environments: High-mass star-forming clumps \citep[black filled circles,][]{Fontani2019}, shock-dominated Galactic Centre clouds \citep[red circles;][]{Rivilla2018}; radiation-dominated Galactic Centre clouds \citep[blue circles;][]{Rivilla2018}, the Orion Bar (blue square, Cuadrado et al. priv. comm.); the L1157 B1 shock \citep[magenta open star;][]{Lefloch2016}. Adapted from \cite{Fontani2019}.}
    \label{fig:PNvsSiO}
\end{figure}

As reported in Section~\ref{subsec:abund}, we do not detect any PN emission above the 3$\sigma$ level and estimate the $\chi$(PN) upper limits, listed in Table~\ref{tab:tab2}. As shown in Figure~\ref{fig:PNvsSiO}, the value obtained for W44 follows the positive trend between $\chi$(PN) and $\chi$(SiO) presented by \cite{Rivilla2018} and \cite{Fontani2019} for different environments such as radiation-dominated Galactic Center clouds, Orion, shock-dominated Galactic Center clouds, protostellar shocks and high-mass clumps. On the other hand, toward IC443, the obtained $\chi$(PN) upper limit is $\sim$1 order of magnitude lower than what reported toward regions with similar SiO molecular abundances.\\

\section{Comparison between models and observations}\label{sec:models}
\begin{figure*}
  \centering
  \includegraphics[width=\columnwidth]{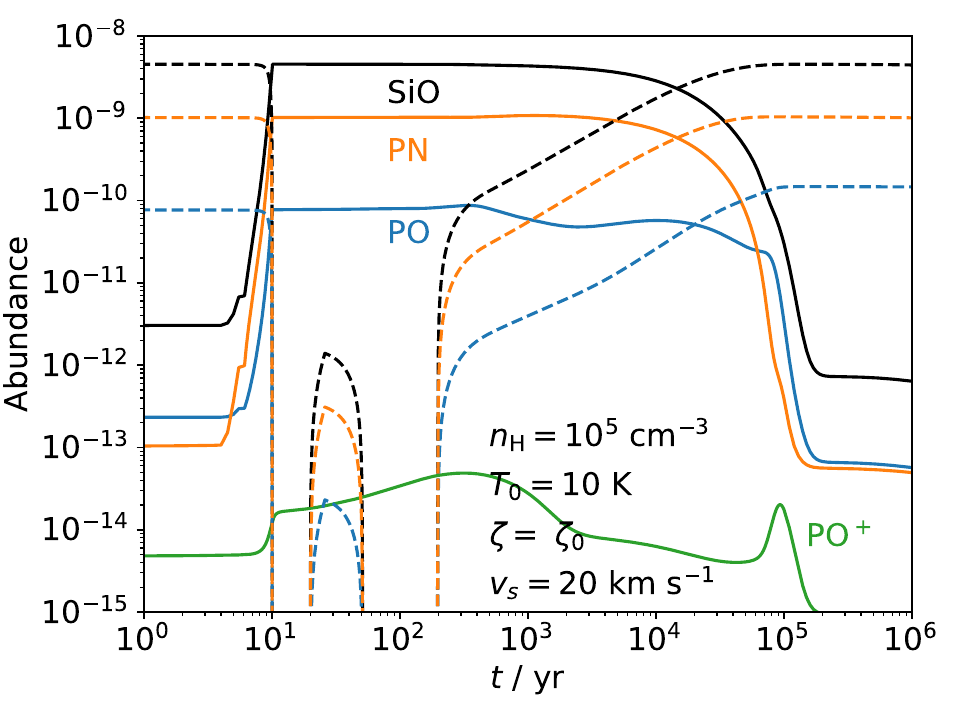}
  \includegraphics[width=\columnwidth]{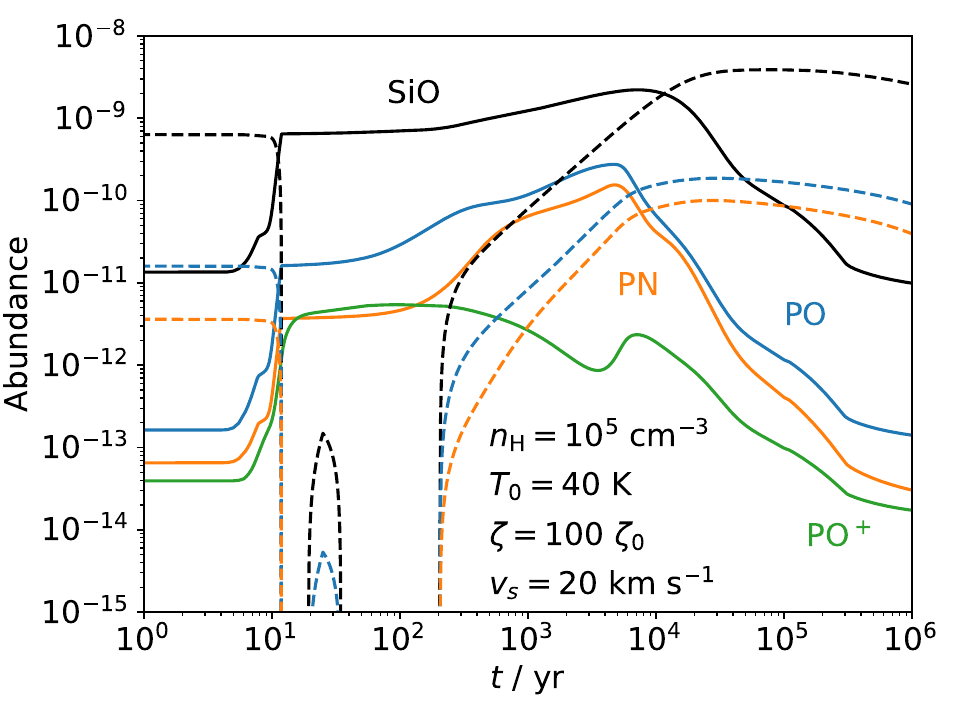}
  \caption{Time evolution of SiO, PO, PN and PO$^+$ molecular abundances in C-type shock models obtained using the UCLCHEM network. In both panels, the used initial parameters are reported. Solid lines represent gas-phase species, dashed lines those in grain mantles.}
  \label{fig:shock}
\end{figure*}

\begin{figure}
  \centering
  \includegraphics[width=\columnwidth]{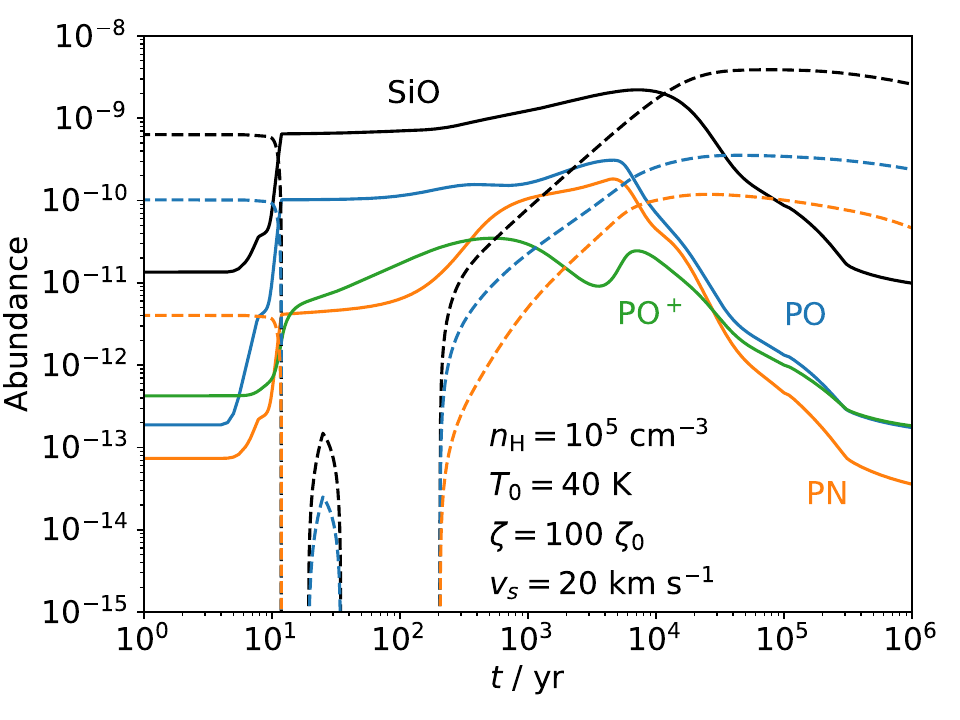}
  \caption{Time evolution of SiO, PO, PN and PO$^+$ molecular abundances in C-type shock models obtained using enhanced cosmic ray ionisation rate and initial gas temperature. A PO$^+$ dissociative recombination rate reduced by a factor of ten was used. Solid lines represent gas-phase species, dashed lines those in grain mantles.}
  \label{fig:lodr}
\end{figure}

From the PN and PO$^+$ abundances in Table~\ref{tab:tab2}, we estimate PO$^+$/PN ratios $\geq$1-5. These values are significantly higher than those reported by \citep{Rivilla2022} and \citep{Scibelli2025} i.e., PO$^+$/PN$\sim$0.04-0.1. We note that our estimates relies on several assumptions and only a multi-transition, high-angular and spectral resolution analysis can fully confirm these values. However, if confirmed, they would suggest an extremely efficient ion-driven phosphorus chemistry in the shocked gas. Hence, we now investigate the gas and shock conditions which might lead to the unusually-high PO$^+$/PN ratio using time-dependent chemical modelling of C-type shocks. We use the astrochemical model {\sc uclchem} \citep{holdship2017} with a two-phase gas-grain network based on UMIST12 \citep{mcelroy2013} which we have extended to include P-bearing species. We consider P, P$^+$, PH, PX, PX$^+$, HPX, and HPX$^+$, where X is any of C, N or O. Charged species are assumed to freeze out as their neutral counterparts, and we adopt binding energies from the KIDA database \citep{wakelam2012}. We do not include PH$_3$, as it has not been detected in molecular cloud environments to date. We use the same elemental abundances as \citet{jimenezserra2018}, specifically assuming a depleted P abundance of $2.57 \times 10^{-9}$. The Si abundance ($8 \times 10^{-9}$) is also strongly depleted, which has previously been found necessary to reproduce observed SiO abundances.

We obtain initial abundances by allowing the chemistry to evolve at fixed physical properties for $10$ Myr, then use the C-type shock parametrization from \citet{jimenez2008} to evolve the gas density and temperature for a further $1$ Myr after the onset of the shock. In the following Figures, only the post-shock time evolution is shown. The main effect of the shock is to release species from ice mantles into the gas phase, where the evolving physical conditions alter their abundances before re-freezing onto grain surfaces after $\sim 10^4$ yr (\textbf{for an initial} medium density of 10$^5$ yr). Rather than a comprehensive study of how changes to the model parameters affect the abundances of P-bearing species as in \citet{jimenezserra2018}, we aim to identify what physical properties are required to produce the PO$^+$/PN ratio of order unity or above, suggested by our observations.

Figure \ref{fig:shock} shows results from C-type shock models with initial density $10^5$ cm$^{-3}$ and shock velocity $20$ km s$^{-1}$, assuming either standard dark cloud conditions for the temperature and cosmic ray ionisation rate ($10$ K, $1.3 \times 10^{-17}$ s$^{-1}$), or enhanced values such as one might expect to find near SNRs ($40$ K, $1.3 \times 10^{-15}$ s$^{-1}$). The temperature value adopted for the enhanced cosmic ray ionisation rate follows the prescription by \cite{cusack2025}. By using hydrodynamic simulations of molecular clouds coupled with an extensive thermochemical network,  a cosmic ionisation rate of ~10$^{15}$ s$^{-1}$ results in temperatures of ~40 K in the 10$^{4}$-10$^{5}$ cm$^{-3}$ density range typical of our sources. The standard conditions fail to reproduce the observed behaviour: PO$^+$ never exceeds PN, and only comes within a factor of a few at late post-shock times, when the gas-phase abundance is significantly lower than observed. With enhanced ionisation rate and initial gas temperature, the predicted PO$^+$/PN ratio is closer to unity both within the first $10-200$ yr after the shock and at $\sim$10$^4$ yr, when the gas phase species onto dust grains are expected to \textbf{freeze-out} onto dust grains, at the considered densities. We note that these results are valid also when higher shocks velocities are considered. Similarly, higher (lower) pre-shocked gas densities mostly have the effect of shortening (prolonging) the post-shock time after which species go back into a depleted stage.

In our model, PO$^+$ forms through a variety of ion-neutral reactions, with different production channels dominating in different regions of the shock depending on physical conditions and reactant abundances. However, the only PO$^+$ destruction mechanism is dissociative recombination with free electrons; neither the UMIST nor KIDA databases list any other reactions which destroy PO$^+$. This reaction is therefore critical for controlling the PO$^+$ abundance throughout the entire shock. The PO$^+$ dissociative recombination rate adopted in both UMIST12 and KIDA was first introduced by \citet{thorne1984} and parametrized as: 
\begin{equation}
k(T) = 3.0 \times 10^{-7}
\left(\frac{T}{300\,\mathrm{K}}\right)^{-0.5}
\ \mathrm{cm^{3}\,s^{-1}} .
\end{equation} 
As already stated by \cite{thorne1984}, the adopted value is an educated guess and no dedicated studies are available to date. However, the abundances of molecular ions in highly-ionised environments such as SNRs can be strongly affected by the adopted dissociative recombination rates \citep[e.g.][]{priestley2017}. We find that already by reducing the \citet{thorne1984} value by a factor of ten,  PO$^+$/PN ratios greater than unity can be produced during the post-shock refreezing period ($\sim$10$^4$ yr), as shown in Figure \ref{fig:lodr}. For comparison, previous shock modelling suggests characteristic interaction timescales larger  than $10^4$ yr for W44 and $10^3$ yr for IC443 \citep[e.g.][]{cosentino2019,reach2019}. However, we note that these value are highly uncertain. Future efforts are needed to obtain a more accurate estimate of this reaction rate in studies of phosphorus chemistry, particularly in extreme environments such as SNR shocks. 

In light of all this, an enhanced cosmic ionisation rates and initial gas temperatures may be necessary to explain the relatively high PO$^+$/PN ratio here reported, as already suggested by \cite{Rivilla2018, Rivilla2022}. These conditions would naturally be expected to occur in the vicinity of an expanding SNR, with the raised gas temperature directly resulting from the increased ionisation rate \citep{cusack2025}. High-angular and spectral resolution studies are needed to confirm the non-detection of PN, assess whether PO is present and to spatially constrain the distribution of these species with respect to SiO.

\section{Discussion}\label{sec:discussion}
SNR shocks are believed to have a strong local impact on the physics and the chemistry of star-forming regions \citep{inoueFukui2018,sano2020,sano2023,khullar2024}. Elements of pre-biotic interest can be released into the gas phase by shock passages and eventually be locked into the material in protoplanetary disks \citep{Fragile2003,lis2016,cosentino2026}. These elements may ultimately play a key role in the emergence of life on planetary systems. Despite this crucial role, only few large molecular surveys toward SNR shocks are present in literature \citep[e.g.][]{vanDishoeck1998,mazumdar2022,tu2025} and even fewer have focused on the presence of COMs and pre-biotic molecules \citep{cosentino2026,shimonishi2026}.\\ 

In this work, we have investigated the presence of the two P-bearing species, PO$^+$ and PN, toward the interaction sites between the SNRs W44 and IC443 and their nearby molecular clouds. We have reported the detection of PO$^+$ with confidence $>$3$\sigma$ toward both SNRs and simultaneously highlighted a lack of PN emission above the detection level. This lack of PN in W44 and IC443 implies PO$^+$/PN ratios $\geq 1$-$5$ i.e., orders of magnitude higher than those reported by \cite{Rivilla2022} and \cite{Scibelli2025}, where PN is more abundant than PO$^+$ by a factor 2-10, but less abundant than PO by a factor $>$3.\\

The non-detection of PN may reflect its efficient destruction in the highly ionised SNR environment. Indeed, recent theoretical work has shown that variations in a few physical parameters, such as the ionisation rate, can significantly alter the balance between neutral and ionised P-bearing species \citep{jimenezserra2018,garcia2021,FernandezRuiz2023,FernandezRuiz2025}. Both W44 \citep[$\sim$10$^{-16}$-$10^{-15}$ s$^{-1}$;][]{cosentino2019} and IC443 \citep[$\sim$10$^{-15}$-$10^{-14}$ s$^{-1}$;][]{Indriolo2010} are known to exhibit elevated cosmic-ray ionisation rates. Under these conditions, PN can be rapidly destroyed through reactions with ions such as H$^+$ and He$^+$ \citep{Chantzos2020}, leading to the formation of P$^+$ and other fragments. These high levels of ionisation may also favour the production of PO$^+$ via direct ionisation of PO and other ion-molecule reactions involving this species \citep{jimenezserra2018}. Since PO is not covered in our frequency bandwidth, additional observations are needed to establish the relative abundance between PO  and PO$^+$ and  whether the obtained PO$^+$ abundances are due to a chemical re-distribution of P-bearing species or an enhancement of Phosphorous in the gas-phase.\\

We speculate that the high PO$^+$/PN ratios here reported may be a direct consequence of the enhanced ionisation in SNR-shocks. Consistent with this scenario, the PN and PO$^+$ abundances measured in this work are better reproduced by astrochemical models that assume cosmic-ray ionisation rates orders of magnitude higher than those typically found in the quiescent ISM. We note that the cosmic ionisation rates considered in the models and previously estimated toward W44 and IC443 are not too far off from those reported by \cite{Rivilla2022} toward the Galactic Center molecular cloud G$+$0.693. However, toward the two SNRs we measure a PO$^+$/PN ratios significantly larger than those reported by \cite{Rivilla2022,Rivilla2018}. This discrepancy suggests that other gas and/or shock properties may also play a role, such as the densities reached by the post-shocked gas and different shock evolutionary stages. A combination of all these parameters may be able to shift the phosphorus chemistry in SNR-impacted gas toward an ion-dominated regime.\\

The PO$^+$ emission toward both W44 and IC443 is associated with strong SiO emission, i.e. with the presence of shocked gas. This is similar to what observed toward the shocked core in the Perseus molecular cloud \cite{Scibelli2025} and the dark cloud G+0.963 toward the Galactic Center \citep{Rivilla2022}. Toward both IC443 and W44, the PO$^+$ emission shows the same number of velocity components of the SiO emission and similar linewidths. This suggests that PO$^+$, as SiO, traces shocked gas. However, systematic velocity offsets are observed between the two tracers. In W44, the PO$^+$ emission is slightly blue-shifted with respect to the SiO emission. In particular, the PO$^+$ centroid velocity is consistent with the low-velocity edge of the gas processed by the J-type component of the shock \cite{cosentino2019}. Toward IC443, the two PO$^+$ velocity components are each one red-shifted with respect to an SiO velocity component. The SiO centroid velocities here identified are consistent with the centroid velocities of the H$_2$ emission reported by \cite{reach2019}. 
In this study, the two velocity components are interpreted as two separate CJ-type shocks ov velocities $\sim$40 and $\sim$60 km s$^{-1}$, plunging into a medium of density $^\sim$10$^3$ cm $^{-3}$ and $\sim$10$^{4}$ cm$^{-3}$, respectively \citep[see also][]{dellova2020,dellova2024,Reach2024}. The presence of multiple velocity components in the PO$^+$ emission toward IC443 may reflect the coexistence of at least two overlapping shocks. Moreover, at least one of the PO$^+$ velocity components is associated with the region of high shock compression reported by \cite{cosentino2022}. Hence, we speculate that PO$^+$ may be preferentially associated with the most compressed and possibly more strongly ionised regions of the shock. However, high-angular and spectral resolution images of the PO$^+$ emission are needed to confirm this hypotheses.\\ 

The comparison between PO$^+$ and SiO abundances presented in Figure~\ref{fig:abund} suggests a possible positive correlation between the two species. However, the number of measurements is too scarce to draw statistically robust conclusions. From Figure~\ref{fig:abund}, while PO$^+$ is enhanced in shocks, its abundance appears to strongly depend on the environment. The PO$^+$ abundance toward W44 is consistent with that measured toward the Galactic Centre cloud G+0.693 \citep{Rivilla2022}, but the SiO abundance is approximately two orders of magnitude lower. On the other hand, in IC443 the SiO abundance is similar to that observed in G+0.693, but the PO$^+$ abundance is higher by at least one order of magnitude. These differences are likely due to a combination of shock strength and different physical conditions across these environments. For instance, not just the shocks driven by IC443 are faster than those driven by W44, but the interacting cloud is regarded to be a translucent cloud \citep{vanDishoeck1998, troja2006,troja2008,cosentino2022}. As a consequence, the H$_2$ column density is significantly lower than that reported toward the dark cloud G034.77-00.55 \citep[$\sim$10$^{22}$ cm$^{-2}$;][]{cosentino2025a} interacting with W44 and the G+0.693 cloud \citep[$\sim$10$^{23}$ cm$^{-2}$;][]{martin2008}. In the lower density environment around IC443, comic rays can reach further depth within the cloud with consequent boost of PO$+$ production. We therefore suggest that the PO$^+$ abundance levels in molecular clouds impacted by SNR feedback may be set by a combination of shock strengths and ionisation power. Expanding the sample of sources with PO$^+$ detections will be crucial to further investigate this potential correlation.\\

Our study suggests that SNR-driven shocks can enhance the presence of P-bearing species in the gas phase of impacted molecular clouds with respect to more quiescent clouds. Current observations do not allow to establish whether this occurs through the injection of phosphorus into the gas phase, or via the processing of refractory material, or through a redistribution of gas-phase phosphorus among different chemical forms. Given the fundamental role of Phosphorus in pre-biotic chemistry, our findings may have important implications for the chemical enrichment of star-forming regions. In the framework of the Sun having formed within a nebula impacted by at least one SNR \citep[e.g.][]{BossKeiser2010}, SNR-driven shocks may have contributed not only to trigger star formation, but also to shape the initial chemical inventory of the protosolar nebula. In particular, we speculate that the highly reactive PO$^+$ released in SNR shocks may represent a transient reservoir of phosphorus that can be recycled into the gas-phase P chemistry, potentially contributing to the formation of more stable species such as PO and PN and, ultimately, to a richer phosphorus chemistry. Alternatively, PO$^+$ may be produced by a chemical re-distribution of P-bearing species, e.g. via ionization of PO. The Phosphorus processed through such shock-driven chemistry may then be incorporated into planetary systems and can be seen, for instance, in the pristine refractory material of comets and meteorites \citep{Rivilla2020}. Future ultra high-sensitivity, high-angular resolution surveys are needed to explore this scenario.

\section{Conclusions}\label{sec:conclusions}
In this paper, we have used IRAM-30m high-sensitivity 3 mm spectra to investigate the presence of PN and PO$^+$ emission at the interaction site between the SNRs W44 and IC443 and their nearby molecular clouds. The obtained results can be summarised as follows:
\begin{itemize}
    \item[i)] We detect PO$^+$ emission above the 3$\sigma$ level toward both SNRs but no significant PN emission. We suggest that the highly ionised SNR environment may shift the chemistry toward one that is dominated by ion-molecule reactions. Such environment favours the production of PO$^+$ over PN. This is consistent with model predictions employing enhanced cosmic ionisation rates and initial gas temperatures. 
    
    \item[ii)] The detected PO$^+$ emission shows linewidth and number of velocity components consistent with the corresponding SiO emission in both sources. However, the PO$^+$ centroid velocities appear blue-shifted with respect to the SiO centroid velocity toward W44 and red-shifted toward IC443. Hence, we speculate that the PO$^+$ emission, contrary to SiO emission, may only trace the highest region of compression enabled by J-type shocks. 
    
    \item[iv)] By comparing the few PO$^+$ abundance estimates available in literature with the corresponding SiO abundance, we speculate that a positive trend between the two quantities may exist. However, the PO$^+$ abundance may be strongly susceptible to a combination of both shock and environmental properties such as shock stage,  post-shocked gas densities, cosmic ionisation rates and gas temperatures. 
    
    \item[v)] The enhanced abundance of PO$^+$ in SNR shocks suggest that these objects may be able to inject material of pre-biotic interest in impacted molecular clouds. PO$^+$ may therefore represent a transient reservoir of Phosphorous from which more complex P-bearing species may form. Hence, SNR-shocks may contribute to the phosphorus enrichment of star-forming environments both by injecting chemically processed material into the surrounding medium and by releasing phosphorus locked in dust grains through shock processing.

\end{itemize}

\begin{acknowledgements}
This work is based on observations carried out under projects with numbers 028-20 and 027-21 with the IRAM 30 m telescope. IRAM is supported by INSU/CNRS (France), MPG (Germany) and IGN (Spain). FDP acknowledges support from the UK Science and Technology Facilities Council Small Award UKRI1187. S.S. acknowledges the National Radio Astronomy Observatory is a facility of the National Science Foundation operated under cooperative agreement by Associated Universities, Inc.
\end{acknowledgements}

\bibliographystyle{aa} 
\bibliography{aa.bib}

\begin{appendix}
\nolinenumbers
\section{Non-LTE analysis}\label{app:radex}
In order to investigate the effects of possible sub-thermal excitation on the column densities derived in Sect.~3.1, we performed non-LTE radiative-transfer calculations for PO$^+$ and PN using RADEX \citep{tak2007} and the collisional rate coefficients available in the LAMDA database \citep{tobola2007,Tonolo2024}. For this, we adopted the plane-parallel geometry in RADEX to approximate the shocked gas layer. Since for each species only one rotational transition is covered by our observations, the gas kinetic temperature, H$_2$ volume density, and molecular column density cannot be simultaneously constrained. Therefore, we used RADEX to explored a grid covering $T_{\rm kin}$=10-100 K and $n({\rm H_2})=10^3-10^6$ cm$^{-3}$ and investigate at which combination of parameter sub-thermal excitation becomes relevant. 

\begin{figure}
    \centering
    \includegraphics[width=\linewidth]{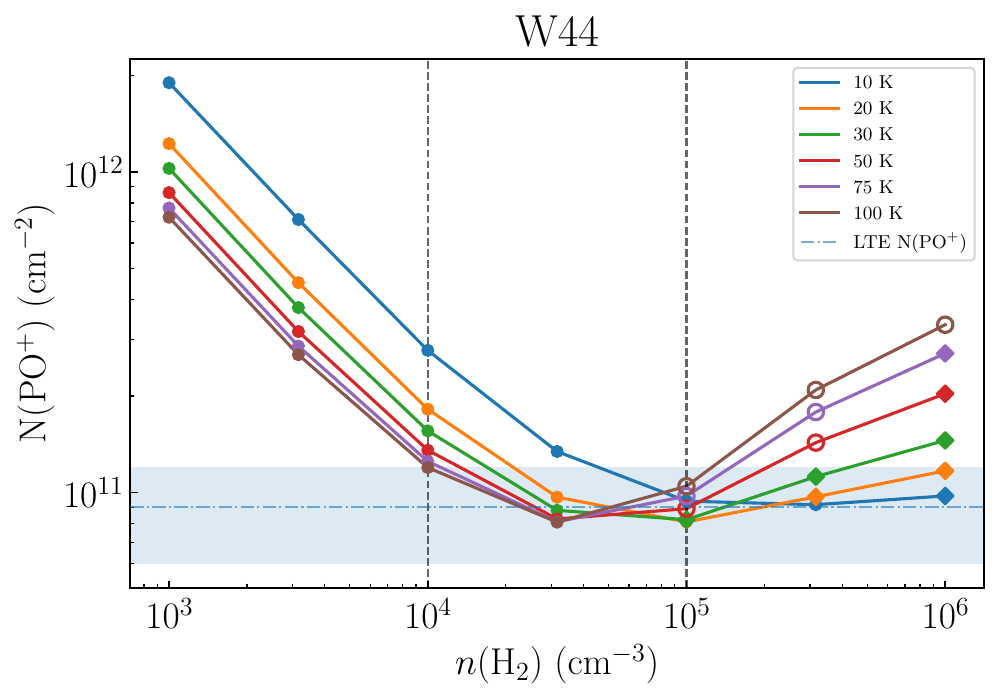}
    \includegraphics[width=\linewidth]{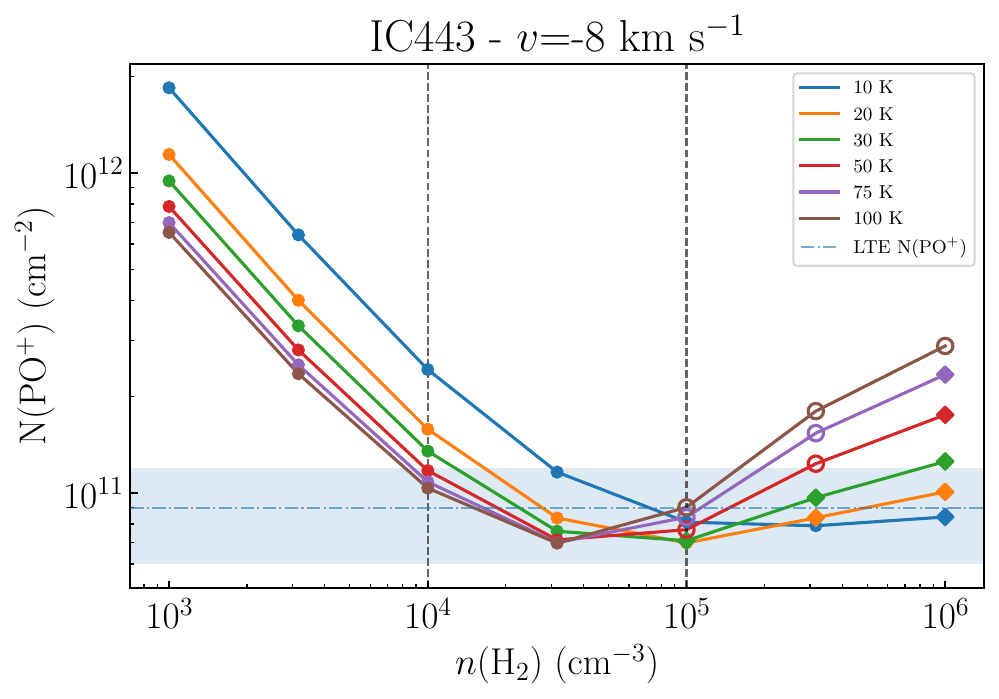}
    \includegraphics[width=\linewidth]{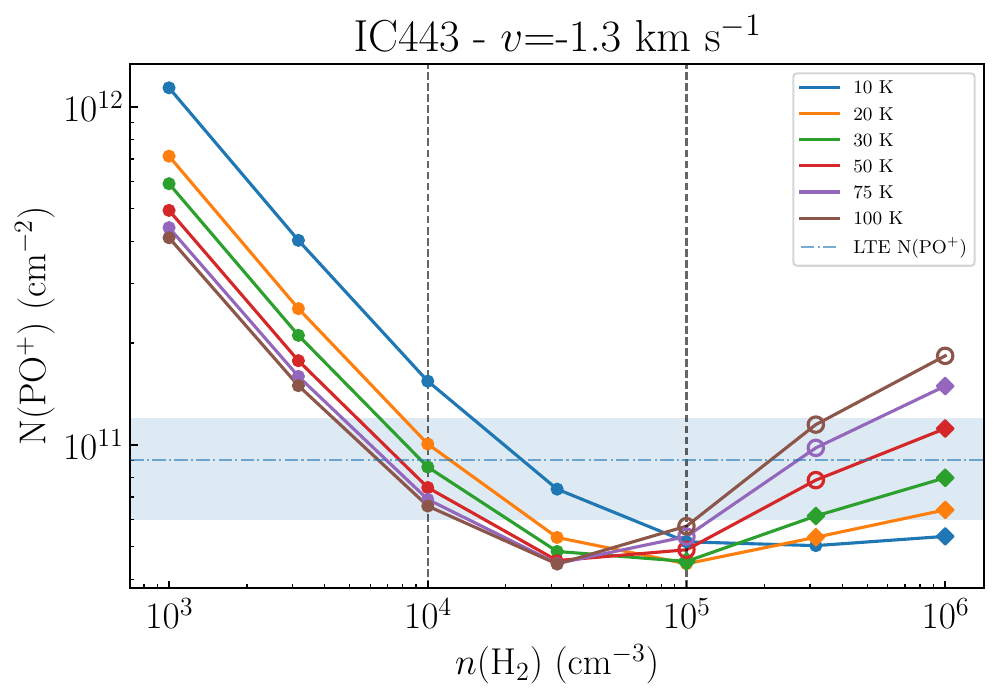}
    \caption{PO$^+$ column densities obtained from the RADEX non-LTE calculations as a function of H$_2$ volume density for W44-Shock (top) and the two velocity components detected toward IC443-P1 (middle and bottom). Different colours correspond to the explored kinetic temperatures. The horizontal dash-dotted lines and shaded regions indicate the column densities derived under the LTE assumption and their uncertainties, respectively. The vertical dashed lines indicate $n({\rm H_2})=10^4$ and $10^5$ cm$^{-3}$, representative of the density range of the compressed gas. Filled circles indicate solutions with $0<T_{\rm ex}\leq T_{\rm kin}$, diamonds indicate supra-thermal excitation ($T_{\rm ex}>T_{\rm kin}$), and open circles indicate solutions showing population inversion ($T_{\rm ex}<0$).}
    \label{fig:radex_po}
\end{figure}

In the case of PO$^+$, for each combination of $T_{\rm kin}$ and $n({\rm H_2})$, we varied $N({\rm PO^+})$ between $10^8$ and $10^{15}$ cm$^{-2}$ and determined the column density that reproduces the observed peak intensity of the PO$^+$(2--1) transition. The linewidth measured for each velocity component was adopted in the calculations, i.e. between 3 and 11 km s$^{-1}$. The resulting $N({\rm PO^+})$ values are shown in Figure~\ref{fig:radex_po} for both W44 and the two velocity component detected toward IC443. 

At $n({\rm H_2})\lesssim10^4$ cm$^{-3}$, the PO$^+$(2--1) transition is strongly sub-thermally excited (filled circles) and column densities significantly larger than the LTE case are required to reproduce the observations. At densities $>10^4$ cm$^{-3}$, the inferred column densities become progressively less dependent on density and approach the values derived under the LTE assumption. We note that, as estimated by \cite{cosentino2019} and \cite{cosentino2022}, the compressed gas in both W44 and IC443 has densities $>n({\rm H_2})\sim10^4$ cm$^{-3}$. In this density regime, representative of the compressed post-shock gas, the plausible kinetic and excitation temperatures are below 30 K and the non-LTE column densities are comparable to the LTE estimates.

As shown in Figure~\ref{fig:radex_po}, for $T_{\rm kin}>20$ K, some solutions at $n({\rm H_2})\gtrsim10^5$ cm$^{-3}$ show supra-thermal excitation (diamond markers), with $T_{\rm ex}>T_{\rm kin}$. Such behaviour can occur for low-$J$ transitions of linear molecules because radiative selection rules couple adjacent rotational levels, while the spontaneous-emission coefficients increase with increasing $J$. The faster radiative decay of the higher-$J$ levels can therefore lead to an overpopulation of the lower rotational levels relative to LTE, resulting in excitation temperatures exceeding the kinetic temperature \citep[e.g.][]{koeppen1980,cosentino2022}. For some of the highest temperatures and densities explored, the calculations formally predict population inversion, characterised by $T_{\rm ex}<0$ (open symbols in Figure~\ref{fig:radex_po}). Given the low observed brightness temperatures of the PO$^+$ emission, we do not consider maser amplification a plausible interpretation of the detected emission. Indeed, as discussed in \cite{cosentino2022}, already the brighter SiO emission observed toward the same regions does not show maser behaviour.

\begin{figure}
    \centering
    \includegraphics[width=\linewidth]{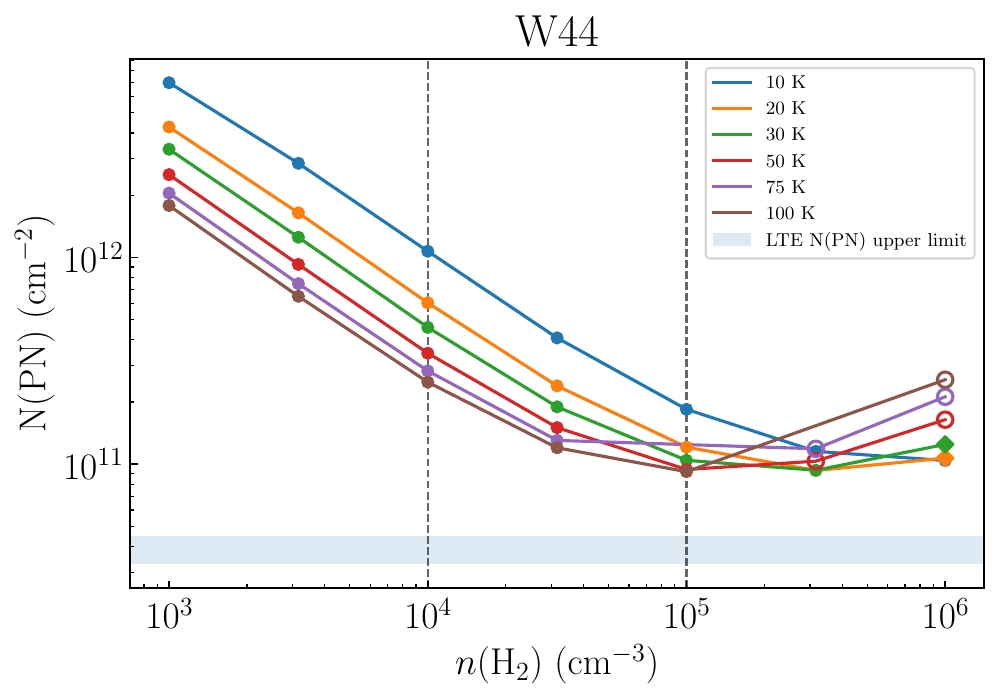}
    \includegraphics[width=\linewidth]{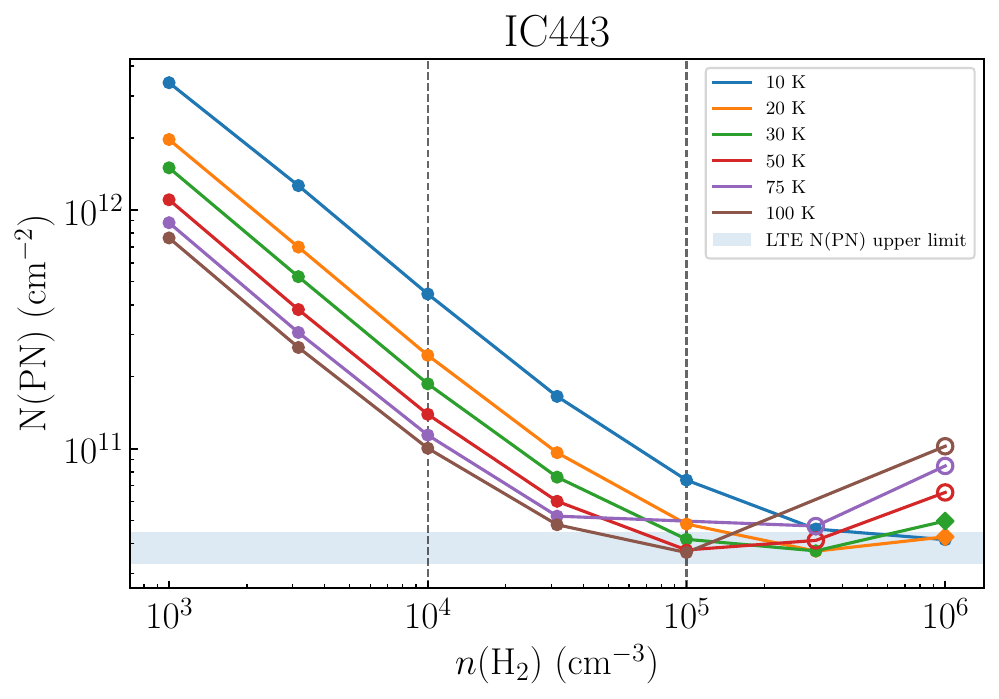}
    \caption{Upper limits on the PN column density obtained from the RADEX non-LTE calculations as a function of H$_2$ volume density toward W44 (top) and IC443 (bottom). For each combination of $T_{\rm kin}$ and $n({\rm H_2})$, $N({\rm PN})$ was varied until the predicted peak intensity of the PN(2--1) transition reached the observational $3\times{\rm rms}$ upper limit. Different colours correspond to the explored kinetic temperatures. The horizontal dash-dotted line and shaded region indicate the range of PN column-density upper limits derived under the LTE assumption toward the two sources. The vertical dashed lines indicate $n({\rm H_2})=10^4$ and $10^5$ cm$^{-3}$. Filled circles indicate solutions with $0<T_{\rm ex}\leq T_{\rm kin}$, diamonds indicate supra-thermal excitation ($T_{\rm ex}>T_{\rm kin}$), and open circles indicate solutions showing population inversion ($T_{\rm ex}<0$).}
    \label{fig:radex_pn}
\end{figure}

For PN, we followed the same procedure, varying $N({\rm PN})$ until the predicted peak intensity of the PN(2--1) transition reached the observational 3$\times$rms upper limit. We adopted a linewidth of 3 km s$^{-1}$, as in the LTE analysis. The results are shown in Figure~\ref{fig:radex_pn}.

PN shows a dependence on density and T$_{\rm kin}$ similar to that observed for PO$^+$. However, toward W44, the PN spectrum has an rms approximately 2.5 times higher than toward IC443. Consequently, its 3$\times\sigma$ intensity upper limit is less stringent, allowing substantially larger N(PN) in the non-LTE solutions. The difference between the two sources should therefore not necessarily be interpreted as an intrinsic difference in PN excitation or abundance, but is at least partly driven by the different observational sensitivities.

Finally, we evaluated the lower limit on $N({\rm PO^+})/N({\rm PN})$ by comparing the PO$^+$ column density and PN upper limit obtained at the same $T_{\rm kin}$ and $n({\rm H_2})$. Toward IC443, the ratio remains larger than unity over most of the physical conditions relevant to the shocked gas. Toward W44, the constraint is weaker, and $N({\rm PO^+})/N({\rm PN})>1$ is obtained only toward the higher densities explored. The high PO$^+$/PN ratios inferred from the LTE analysis should therefore not be regarded as independent of the assumed excitation conditions, particularly toward W44. Nevertheless, for densities representative of the compressed post-shock gas, the PO$^+$ column densities themselves remain comparable to the LTE estimates.

\begin{figure}
    \centering
    \includegraphics[width=\linewidth]{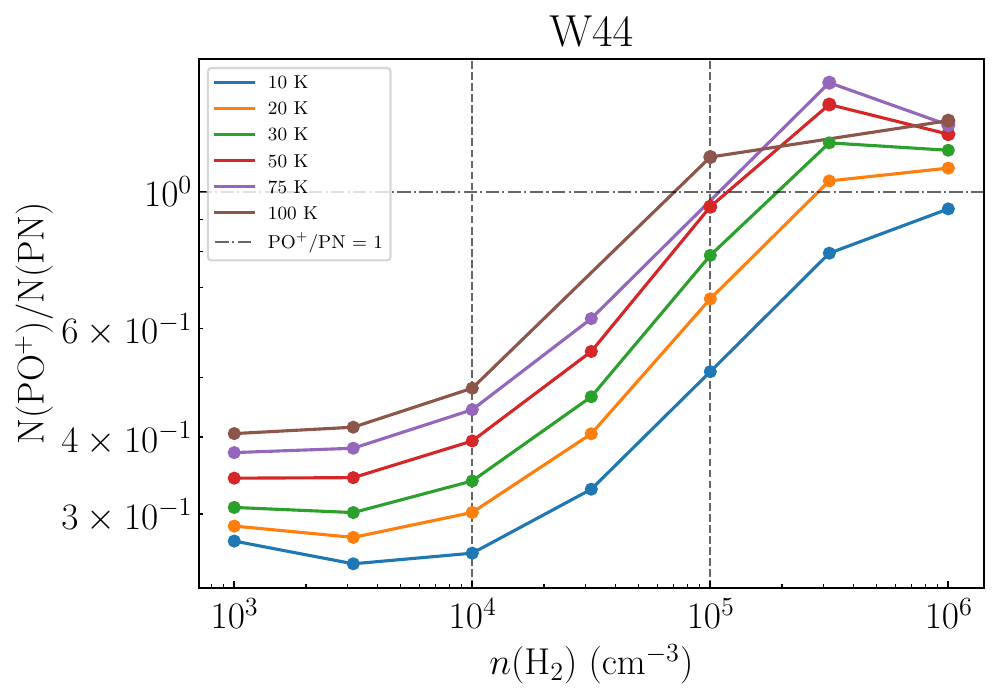}
    \includegraphics[width=\linewidth]{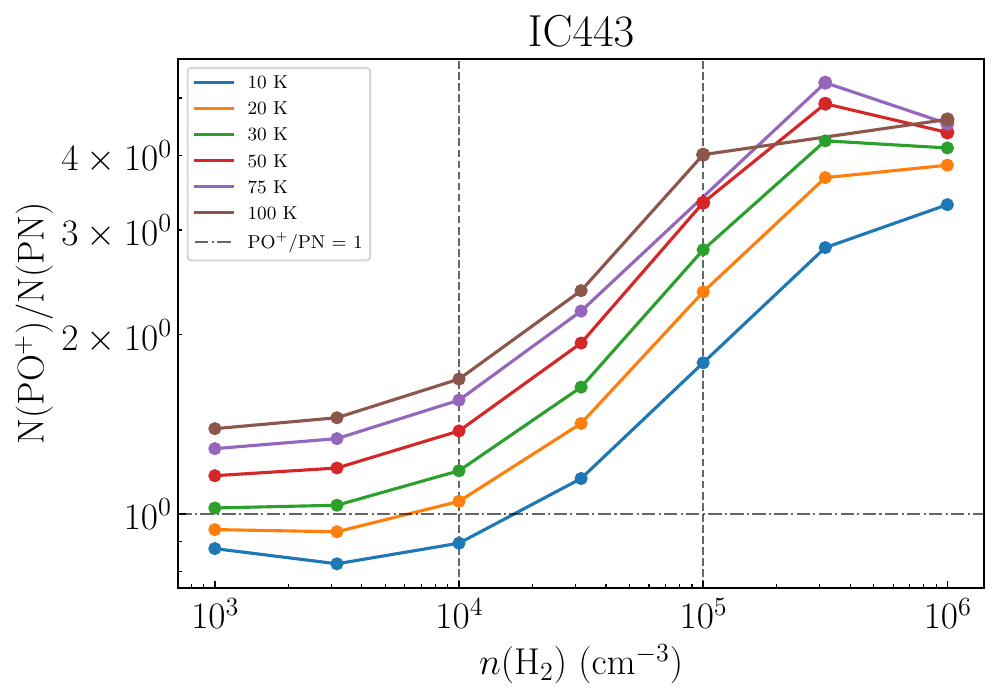}
    \caption{Lower limits on the $N({\rm PO^+})/N({\rm PN})$ column-density ratio obtained from the RADEX non-LTE calculations as a function of H$_2$ volume density toward W44 (top) and IC443 (bottom). The PO$^+$ column densities and PN upper limits were evaluated at the same $T_{\rm kin}$ and $n({\rm H_2})$. Different colours correspond to the explored kinetic temperatures. The horizontal dashed line marks $N({\rm PO^+})/N({\rm PN})=1$, while the vertical dashed lines indicate $n({\rm H_2})=10^4$ and $10^5$ cm$^{-3}$.}
    \label{fig:radex_ratios}
\end{figure}

\end{appendix}
\end{document}